\documentclass[useAMS,usenatbib,usegraphicx]{mn2e} 
\usepackage{amssymb} 
\usepackage{times} 
\usepackage{aas_macros} %needed to define journal macros generated by 
\usepackage[usenames]{color}

\title[FM stars]{FM stars: A Fourier view of pulsating binary stars, a new 
technique for measuring radial velocities photometrically}

\author[Shibahashi \& Kurtz] 
{Hiromoto Shibahashi$^1$ and Donald W. Kurtz$^2$ \\ 
$^{1}$Department of Astronomy, The University of Tokyo, Tokyo 113-0033, Japan \\
$^{2}$Jeremiah Horrocks Institute, University of Central 
Lancashire, Preston PR1 2HE, UK\\ }
 
\begin{document} 

\maketitle 

\begin{abstract} 
Some pulsating stars are good clocks. When they are found in binary stars, the 
frequencies of their luminosity variations are modulated by the Doppler effect 
caused by orbital motion. For each pulsation frequency this manifests itself as a 
multiplet separated by the orbital frequency in the Fourier transform of the light 
curve of the star. We derive the theoretical relations to exploit data from the 
Fourier transform to derive all the parameters of a binary system traditionally 
extracted from spectroscopic radial velocities, including the mass function which 
is easily derived from the amplitude ratio of the first orbital sidelobes to the 
central frequency for each pulsation frequency. This is a new technique that 
yields radial velocities from the Doppler shift of a pulsation frequency, thus 
eliminates the need to obtain spectra. For binary stars with pulsating components, 
an orbital solution can be obtained from the light curve alone. We give a complete 
derivation of this and demonstrate it both with artificial data, and with a case 
of a hierarchical eclipsing binary with {\it Kepler} mission data, KIC\,4150611 
(HD\,181469). We show that it is possible to detect Jupiter-mass planets orbiting 
$\delta$\,Sct and other pulsating stars with our technique. We also show how to 
distinguish orbital frequency multiplets from potentially similar nonradial 
$m$-mode multiplets and from oblique pulsation multiplets.
\end{abstract} 


%%%%%%%%%%%%%%%%%%%%%%%%%%%%%%%%%%%%%%
\section{Introduction} 
\label{sec:1}

There are many periodic phenomena in astronomy that act as clocks: the Earth's 
rotation and orbit, the orbit of the Moon, the orbits of binary stars and 
exoplanets, the spin of pulsars, stellar rotation and stellar pulsation are 
examples. All of these astronomical clocks show measurable frequency and/or phase 
modulation in the modern era of attosecond precision atomic time, and stunning 
geophysical and astrophysical insight can be gleaned from their frequency 
variability.

Variations in the Earth's rotation arise from, for example, changes in seasonal 
winds, longer-term changes in ocean currents (such as in the El Nino 
quasi-biennial southern oscillation), monthly changes in the tides, 
and long-term tidal 
interaction between the Earth and Moon. Earth rotation even suffers measurable 
glitches with large earthquakes and internal changes in Earth's rotational angular 
momentum. Earth's rotation and orbit induce frequency variability in all 
astronomical observations and must be precisely accounted for, usually by 
transforming times of observations to Barycentric Julian Date (BJD). The 
astronomical unit is known to an accuracy of less than 5\,m, on which scale the 
Earth's orbit is not closed and is highly non-Keplerian. Incorrectly transforming 
to BJD led to the first claim of discovery of an exoplanet 
(\citealt{bailesetal1991}; it was a rediscovery of Earth), and correct 
transformation led to the true first exoplanets discovered orbiting the pulsar 
PSR\,1257+12 \citep{wolszczanfrail1992}. \citet{hulsetaylor1975} famously 
discovered the first binary pulsar. Timing variations in its pulses confirmed 
energy losses caused by gravitational radiation and led to the award of the Nobel 
prize to them in 1993. 

Deviations of astronomical clocks from perfect time keepers have traditionally 
been studied using `Observed minus Corrected' ($O-C$) diagrams (see, e.g., 
\citealt{sterken2005} and other papers in those proceedings). Pulsar timings are 
all studied this way. In an $O-C$ diagram some measure of periodicity (pulse 
timing in a pulsar, time of periastron passage in a binary star, the phase of one 
pulsation cycle in a pulsating star) is compared to a hypothetical perfect clock 
with an assumed period. Deviations from linearity in the $O-C$ diagram can then 
diagnose evolutionary changes in an orbit or stellar pulsation, apsidal motion in 
a binary star, stochastic or cyclic variations in stellar pulsation, and, most 
importantly for our purposes here, periodic Doppler variability in a binary star 
or exoplanet system. 

That some pulsating stars are sufficiently good clocks to detect exoplanets has 
been demonstrated. In the case of V391\,Peg \citep{silvottietal2007},  a 3.2-
M$_{\rm Jupiter}/\sin i$ planet in a 3.2-yr orbit around an Extreme Horizontal 
Branch star was detected in sinusoidal frequency variations inferred from the $O-
C$ diagrams for two independent pulsation modes (where M$_{\rm Jupiter}$ is the 
Jovian mass and $i$ denotes the inclination angle of the orbital axis with respect 
to the line of sight). These frequency variations arise simply as a result of the 
light time effect. 

A barrier to the study of binary star orbits and exoplanet orbits using $O-C$ 
diagrams has always been the difficulty of precisely phasing cycles across the 
inevitable gaps in ground-based observations. Great care must be taken not to lose 
cycle counts across the gaps. The light time effect can also be seen directly in 
the Fourier transform of a light curve of a pulsating star, where the cycle count 
ambiguity manifests itself in the aliases in the spectral window pattern. In 
principle, this method yields all the information in an $O-C$ diagram, but the 
tradition has been to use $O-C$ diagrams rather than amplitude spectra, probably 
because of the apparently daunting confusion of multiple spectral window patterns 
for multiperiodic pulsating stars with large gaps in their light curves. 

Now, with spaced-based light curves of stars at $\mu$mag photometric precision and 
with duty cycles exceeding 90\,per\,cent, e.g., those obtained with the {\it 
Kepler} mission \citep{kochetal2010}, not only is there no need to resort to $O-C$ 
diagrams to study orbital motion from the light curve, it is preferable to do this 
directly with information about the frequencies derived from the Fourier 
transform. For pulsating stars that are sufficiently good clocks, it is possible 
to derive orbital radial velocities in a binary system from the light curve alone 
-- obviating the need for time-consuming spectroscopic observations. The 
fundamental mass function, 
$f(m_1, m_2, \sin i)= {m_2^3 \sin^3 i}/{(m_1 + m_2)^2}$, 
for a binary star can be derived directly from the amplitudes and phases of 
frequency multiplets found in the amplitude spectrum without need of radial 
velocities, although those, too, can be determined from the photometric data. 
Here, $m_1$ and $m_2$ denote the mass of the pulsating star in the binary system 
and the mass of the companion, respectively, and $i$ is the inclination angle of 
the orbital axis with respect to the line of sight.  

The {\it Kepler} mission is observing about 150\,000 stars nearly continuously for 
spans of months to years. Many of these stars are classical pulsating variables, 
some of which are in binary or multiple star systems. The study of the pulsations 
in such stars is traditionally done using Fourier transforms, and it is the 
patterns in the frequencies that lead to astrophysical inference; see, for 
example, the fundamental textbooks \citet{unnoetal1989} and \citet{aertsetal2010}. 
One type of pattern that arises is the frequency multiplet. This may be the result 
of nonradial modes of degree $\ell$ for which all, or some, of the ($2\ell+1$) 
$m$-mode components (where $-\ell \le m \le +\ell$) may be present. 
The splitting between the 
frequencies of such multiplets is proportional to stellar angular frequency, hence 
leads to a direct measure of the rotation velocity of the star averaged over the 
pulsation cavity \citep{ledoux1951}. In the best case of the Sun, this leads to a 
2D map (in depth and latitude) of rotation velocity over the outer 
half of the solar radius. Frequency multiplets also occur for a star that has 
pulsation modes inclined to the rotation axis, leading to oblique pulsation, as in 
the roAp stars (e.g., \citealt{kurtz1982}; \citealt{shibahashitakata1993}; 
\citealt{bigotkurtz2011}). This, too, leads to a frequency multiplet, in this case 
split by exactly the rotation frequency of the star. 

Other types of frequency modulation may be present in pulsating stars. The Sun is 
known to show frequency variability correlated with the 11-yr solar cycle. A large 
fraction of RR\,Lyr stars exhibit quasi-periodic amplitude and phase modulation 
known as the Blazhko effect. The physical cause of this remains a mystery after 
more than a century of study. \citet{benkoetal2009} and 
\citet{benkoetal2011} have looked at 
the formalism of the combination of frequency modulation and amplitude modulation 
in the context of the Blazhko stars, showing the type of frequency multiplets 
expected compared to those observed. Their frequency modulation is analogous to 
that of FM (frequency modulation) radio waves, hence the formalism is well known 
in the theory of radio engineering, but unfamiliar to most astronomers. 

\begin{figure}
\centering
\includegraphics[width=\linewidth,angle=0]{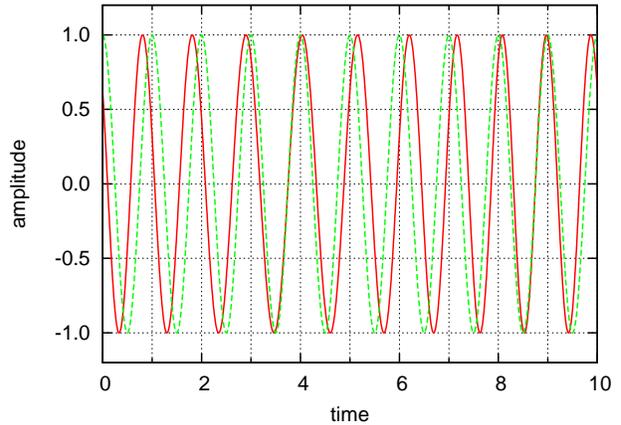}
\caption{A schematic picture of phase modulation. The green, dashed curve shows a 
pure sinusoid. The red, solid curve shows the same sinusoid with frequency 
modulation.}
\label{fig:1}
\end{figure}

Here we examine FM for pulsating stars in binary star systems. Imagine that one 
star in a binary system is sinusoidally pulsating with a single frequency. Its 
luminosity varies with time as a consequence of pulsation. For a single star with 
no radial velocity with respect to the solar system barycentre, the observed 
luminosity variation would also be purely sinusoidal and would be expressed in 
terms of the exact same frequency as the one with which the star is intrinsically 
pulsating. But, in the case of a binary system, the orbital motion of the star 
leads a periodic variation in the distance between us and the star; that is, the 
path length of the light, thus the phase of the observed luminosity variation also 
varies with the orbital period. This is the light time effect and is equivalent to 
a periodic Doppler shift of the pulsation frequency. The situation is the same as 
the case of a binary pulsar. Fig.\,\ref{fig:1} shows the difference between the 
light curves in these two cases. 

In the following section we show the formal derivations of the light time effect 
in a binary star on the Fourier transform of the light curve of a pulsating star. 
This also leads to frequency multiplets in the amplitude spectra of such stars 
where the frequency splitting is equal to the orbital frequency, and where the 
amplitudes and phases of the components of the frequency multiplet can be used to 
derive all of the information traditionally found from radial velocity curves: the 
time of periastron passage, orbital eccentricity, the mass function, $f(m_1,m_2, 
\sin i)$, and even the radial velocity curve 
itself. This is a significant advance in the study of binary star orbits; 
effectively we have photometric radial velocities. 

Asteroseismologists must also be aware of the expected frequency patterns for 
pulsating stars in binary systems. These show a new kind of multiplet in the 
amplitude spectra that needs to be recognised and exploited. In the common case of 
binary stars with short orbital periods where rotation is synchronous, the 
frequency multiplets that we derive here, Ledoux rotational spitting multiplets 
and oblique pulsation multiplets can potentially be confused, and must be 
distinguished. We show how this is possible using frequency separation, amplitude 
ratios and multiplet phase relationships. 

%%%%%%%%%%%%%%%%%%%%%%%%%%%%%%%%%%%%%%
\section{The simplest case: a binary star with circular orbital motion}
\label{sec:2}
\subsection{Analytical expression of phase modulation}
\label{sec:2.1}

Let us first consider the simplest case, a pulsating star in a binary with 
circular orbital motion. We assume that the radial velocity of the centre-of-mass 
of the binary system with respect to the solar system barycentre -- the $\gamma$-
velocity -- has been subtracted, and we assume that observations are corrected to 
the solar system barycentre so that there is no component of the Earth's orbital 
velocity. We name the stars `1' and `2', and suppose that star 1 is pulsating. 
In this case, the observed luminosity variation at time $t$ has a form
\begin{eqnarray}
	\cos\left\{ \omega_0 
	\left[t - {{1}\over{c}}\int_0^t v_{\rm rad, 1}(t') dt' \right] + \phi 
	\right\},
\label{eq:1}
\end{eqnarray}
where $\omega_0$ is the angular frequency of pulsation, $c$ the speed of light, 
$v_{\rm rad, 1}$ denotes the line of sight velocity of the star 1 due to orbital 
motion, and $\phi$ is the pulsation phase at $t=0$. The second term in the 
square bracket measures the arrival time delay of the signal from the star to us. 
The instantaneously observed frequency is regarded as the time derivative of the 
phase, which is given by 
\begin{eqnarray}
	\omega_{\rm obs} 
	= \omega_0 \left[ 1 - {{v_{\rm rad, 1}(t)}\over{c}} \right]. 
\label{eq:2}
\end{eqnarray}
The second term in the right-hand-side of the above equation 
is the classical Doppler shift of the frequency.

In this section, we adopt the phase at which the radial velocity of star 1 reaches 
its maximum, i.e. the maximum velocity of recession, as $t=0$. Then, the radial 
velocity $v_{\rm rad, 1}(t)$ is given by
\begin{eqnarray}
	v_{\rm rad, 1}(t) = a_1 \Omega \sin i \cos\Omega t,
\label{eq:3}
\end{eqnarray}
where $a_1$ denotes the orbital radius of star 1, that is, the distance from the 
star to the centre of gravity of the binary system, $\Omega$ denotes the orbital 
angular frequency, and $i$ denotes the inclination angle of the orbital axis with 
respect to the line of sight. Following convention, the sign of $v_{\rm rad, 1}$ 
is defined so that $v_{\rm rad, 1} > 0$ when the object is receding from us. 
Hence,
\begin{eqnarray}
	&&
	\cos\left\{ \omega_0 
	\left[ t - {{1}\over{c}}\int_0^t v_{\rm rad, 1}(t') dt' \right] + \phi 
	\right\}
	\nonumber \\
	&&
	=
	\cos\left[ \left( \omega_0 t + \phi \right) 
	+ {{a_1\omega_0\sin i}\over{c}}\sin\Omega t \right].
\label{eq:4}
\end{eqnarray}
This expression means that the phase is modulated with the orbital angular frequency 
$\Omega$ and with the amplitude $a_1\omega_0\sin i/c$. This result is reasonable, 
since the maximum arrival time delay is $a_1\sin i/c$, hence the maximum phase 
difference is $a_1\omega_0\sin i/c$.

\subsection{An estimate of the amplitude of phase modulation}
\label{sec:2.2}

From Kepler's 3rd law, the separation between the components 1 and 2 of a binary 
is 
\begin{eqnarray}
	a
	=
	\left({{G{\rm M}_\odot}\over{4\pi^2}}\right)^{1/3} 
	\left({{m_1}\over{{\rm M}_\odot}}\right)^{1/3}(1+q)^{1/3} P_{\rm orb}^{2/3} ,
\label{eq:5}
\end{eqnarray} 
where 
\begin{equation}
	q \equiv {{m_2}\over{m_1}}
\label{eq:6}
\end{equation}
denotes the mass ratio of the stars and 
\begin{equation}
	P_{\rm orb}\equiv {{2\pi}\over{\Omega}}
\label{eq:7}
\end{equation}
denotes the orbital period.
Hence, the distance between star 1 and the centre of gravity, $a_1$, is 
\begin{eqnarray}
	a_1 
	&=&
	q(1+q)^{-1}a
	\nonumber \\
	&=&
	\left({{G{\rm M}_\odot}\over{4\pi^2}}\right)^{1/3} 
	\left({{m_1}\over{{\rm M}_\odot}}\right)^{1/3} q(1+q)^{-2/3} P_{\rm 
orb}^{2/3}.
\label{eq:8}
\end{eqnarray}
Then, the amplitude of the Doppler frequency shift, $a_1\Omega\sin i/c$, is given 
by 
\begin{eqnarray}
	& &
	{{\left(2\pi G{\rm M}_\odot\right)^{1/3} }\over{c}}
	\left({{m_1}\over{{\rm M}_\odot}}\right)^{1/3} q(1+q)^{-2/3} {P_{\rm orb}}^{-
1/3} 
	\sin i 
\nonumber\\
	&\simeq&
	7.1\times 10^{-4}
	\left({{m_1}\over{{\rm M}_\odot}}\right)^{1/3} q(1+q)^{-2/3} 
	\left({{P_{\rm orb}}\over{1\,{\rm d}}}\right)^{-1/3} \sin i .
\nonumber\\
& &
\label{eq:9}
\end{eqnarray}
This is typically of the order of $10^{-3}$. 
 
\begin{figure}
\centering
\includegraphics[width=\linewidth,angle=0]{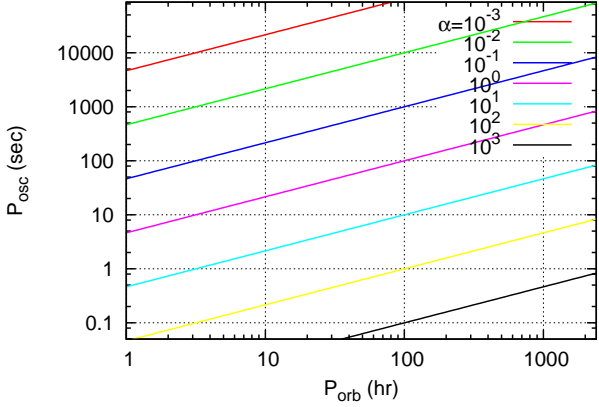}
\caption{The dependence of the phase modulation amplitude $\alpha$ in the case of 
$m_1 = 1$\,M$_{\odot}$, $q=1$, and $i=90^\circ$.}
\label{fig:2}
\end{figure}

The amplitude of phase modulation, $a_1\omega_0\sin i/c$, is given by 
\begin{eqnarray}
	\alpha
	&\equiv&
	{{\left(2\pi G{\rm M}_\odot\right)^{1/3} }\over{c}}
	\left({{m_1}\over{{\rm M}_\odot}}\right)^{1/3} q(1+q)^{-2/3} 
	{{P_{\rm orb}^{2/3}}\over{P_{\rm osc}}} \sin i
\nonumber\\
	& &
\label{eq:10}
	\\
	&\simeq&
	1.7 \times 10^{-2} 
	\left({{m_1}\over{{\rm M}_\odot}}\right)^{1/3} q(1+q)^{-2/3}  
	{{(P_{\rm orb}/1\,{\rm d})^{2/3}}\over{(P_{\rm osc}/1\,{\rm h})}} \sin i,
\nonumber\\
& &
\label{eq:11}
\end{eqnarray}
where
\begin{equation}
	P_{\rm osc}\equiv {{2\pi}\over{\omega_0}}
\label{eq:12}
\end{equation}
denotes the pulsation period. It should be noted that the amplitude of the phase 
modulation is not necessarily small; it can be quite large depending on the 
combination of $P_{\rm orb}$ and $P_{\rm osc}$. In the case of $P_{\rm orb} = 1$\, 
d and $P_{\rm osc} = 1$\,h, the amplitude of phase modulation is of order of 
$10^{-2}$. Fig.\,\ref{fig:2} shows the dependence of the phase modulation 
amplitude, $\alpha$, on $P_{\rm orb}$ and $P_{\rm osc}$ in the case of $m_1 = 
1$\,M$_{\odot}$, $q=1$ and $i = 90^\circ$. 

Larger values of the phase modulation, $\alpha$, are more detectable. It can 
be seen in Fig.\,\ref{fig:2} that for a given oscillation period, longer orbital 
periods are more detectable, and for a given orbital period shorter pulsation 
periods (higher pulsation frequencies) are more detectable. The combination of the 
two -- high pulsation frequency and long orbital period -- gives the most 
detectable cases. This will be relevant in Section\,\ref{sec:4} below when we 
discuss the 
detectability of exoplanets with our technique.

\subsection{Mathematical formulae}
\label{sec:2.3}

\begin{figure}
\centering
\includegraphics[width=\linewidth,angle=0]{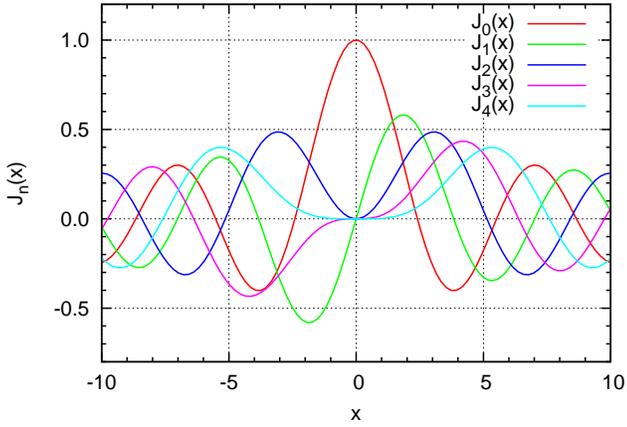}
\caption{Bessel functions of the first kind with integer order.}
\label{fig:3}
\end{figure}

Our aim is to carry out the Fourier analysis of pulsating stars showing phase 
modulation due to orbital motion in a binary. As deduced from 
equation\,(\ref{eq:4}), the problem is then essentially how to treat the terms 
$\cos(\alpha\sin\Omega t)$ and $\sin(\alpha\sin\Omega t)$. These terms can be 
expressed with a series expansion in terms of Bessel functions of the first kind 
with integer order:
\begin{eqnarray}
	\cos(\alpha\sin\Omega t)
	=
	J_0(\alpha) + 2\sum_{n=1}^\infty J_{2n}(\alpha) \cos 2n\Omega t
\label{eq:13}
\end{eqnarray}
\begin{eqnarray}
	\sin(\alpha\sin\Omega t)
	=
	2 \sum_{n=0}^\infty J_{2n+1}(\alpha) \sin (2n+1)\Omega t ,
\label{eq:14}
\end{eqnarray}
respectively\footnote{These relations can be derived from the generating function 
of the Bessel functions $\exp{{1}\over{2}}x(t-t^{-1})$ by replacing $x$ and $t$ with 
$\alpha$ and $\pm\exp(i\Omega t)$, respectively.}. Here $J_n(x)$ denotes the Bessel function of the 
first kind\footnote{According to \citet{watson1922}, the Bessel function of order 
zero was first described by \citet{bernoulli1738}.} of integer order $n$:
\begin{eqnarray}
	J_n(x)=\sum_{k=0}^\infty (-1)^k {{(x/2)^{n+2k}}\over{(n+k)!\,k!}}.
\label{eq:15}
\end{eqnarray}
Fig.\,\ref{fig:3} illustrates the five lowest-order such functions.

Noting that Bessel functions with negative integer orders are defined as 
\begin{eqnarray}
	J_{-n}(x) = (-1)^n J_n(x) ,
\label{eq:16}
\end{eqnarray}
we reach an expression of the right-hand side of equation\,({\ref{eq:4}}) with a  
series expansion in terms of cosine functions:
\begin{eqnarray}
	&&
	\cos \left[ (\omega_0 t + \phi) + \alpha \sin\Omega t \right]
	\nonumber\\
	&&
	=
	\sum_{n=-\infty}^\infty J_n(\alpha) 
	\cos \left[ (\omega_0 + n\Omega) t + \phi \right] .
\label{eq:17}
\end{eqnarray}
It should be noted that this relation is the mathematical base for the broadcast 
of FM radio. This relation also has been applied recently to Blazhko RR\,Lyr stars 
by \citet{benkoetal2009} and \citet{benkoetal2011}. Vibrato in music is another 
example that is described by this equation.

It is instructive to write down here, for later use, similar expansions for 
$\cos(\alpha\cos \Omega t)$ and $\sin(\alpha\cos\Omega t)$ as well:
\begin{eqnarray}
	\cos(\alpha\cos\Omega t)
	=
	J_0(\alpha) + 2\sum_{n=1}^\infty (-1)^n J_{2n}(\alpha) \cos 2n\Omega t
\label{eq:18}
\end{eqnarray}
and
\begin{eqnarray}
	\sin(\alpha\cos\Omega t)
	=
	2\sum_{n=0}^\infty (-1)^n J_{2n+1}(\alpha) \cos (2n+1)\Omega t.
\label{eq:19}
\end{eqnarray}
After the names of two great mathematicians Carl Jacobi and Carl Theodor Anger who 
derived these series expansions, the expansions (\ref{eq:13}), (\ref{eq:14}), 
(\ref{eq:18}), and (\ref{eq:19}) are now called Jacobi-Anger 
expansions\footnote{According to \cite{watson1922}, equations\,(\ref{eq:18}) and 
(\ref{eq:19}) were given by \citet{jacobi1836}, and equations\,(\ref{eq:13}) and 
(\ref{eq:14}) were obtained later by \citet{anger1855}.}.

\subsection{The expected frequency spectrum}
\label{sec:2.4}
\subsubsection{General description}
\label{sec:2.4.1}

\begin{figure}
\centering
\includegraphics[width=\linewidth,angle=0]{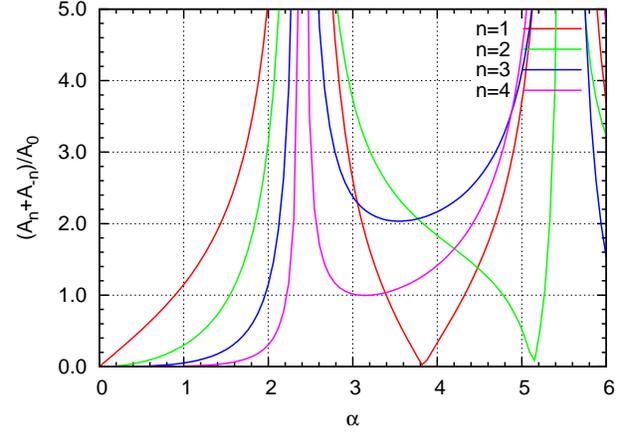}
\caption{The amplitude ratio of the $n$-th side peaks to the central peak of the 
multiplet frequency spectrum, as a function of the phase modulation amplitude 
$\alpha$. $A_{\pm n}$ and $A_0$ denote the amplitude of the frequencies at 
$\omega_0 \pm n\Omega$ and $\omega_0$, respectively. }
\label{fig:4}
\end{figure}

Equation\,(\ref{eq:17}) means that a frequency multiplet equally split by the 
orbital frequency $\Omega$ appears in the frequency spectrum of a pulsating star 
in a binary system with a circular orbit. The orbital period is then determined 
from the spacing of the multiplet. The amplitude ratio of the $n$-th side 
peak to the central peak is given by
\begin{eqnarray}
	{{A_{+n} + A_{-n}}\over{A_0}} = {{2 | J_n(\alpha) |}\over{| J_0(\alpha) |}},
\label{eq:20}
\end{eqnarray}
where $A_{+n}$, $A_{-n}$ and $A_{0}$ represent the amplitudes of the peaks at 
$\omega_0 + n\Omega$, $\omega_0 - n\Omega$, and $\omega_0$, respectively. 
Fig.\,\ref{fig:4} shows the amplitude ratio of the $n$-th peak to the central peak 
as a function of the phase modulation amplitude $\alpha$.

The multiplet is an infinite fold, but the dominant peaks are highly dependent on 
the value of $\alpha$. For example, in the case of an 0.5-M$_{\odot}$ sdB star 
pulsating with $P_{\rm osc}=150\,{\rm s}$ and orbiting with $P_{\rm orb} = 
0.1\,{\rm d}$ with the same mass companion, $\alpha \sim 0.04$ and we expect 
$J_0({\alpha})\simeq 1$, $J_1({\alpha})\sim \alpha/2 \sim 2\times 10^{-2}$, and 
$J_n(\alpha)\sim 0$ for $n \geq 2$. In this case we expect a triplet structure, 
for which the central component, corresponding to $n=0$ with a frequency 
$\omega_0$, is the highest, and the side peaks, separated from the central peak by 
$\Omega$, are of the order of $2\times 10^{-2}$ the amplitude of the central peak. 
However, if the same star is orbiting with $P_{\rm orb} = 12.5\,{\rm d}$, then 
$\alpha \simeq 1$. In this case, $J_1({\alpha})/J_0({\alpha}) \sim 1$ and 
$J_2({\alpha})/J_0({\alpha}) \sim 0.3$ 
so that $(A_{+1} + A_{-1})/{A_0} \sim 2$ and 
$(A_{+2} + A_{-2})/{A_0} \sim 0.6$; 
the contribution of $J_2(\alpha)$ is not 
negligible, and a quintuplet structure with an equal spacing 
of $\Omega$ is expected. 
As seen in Fig.\,\ref{fig:4}, for $\alpha \gtrsim 0.9$ the side peaks could have 
higher amplitude than the central component. 

Importantly, the phases of the components of the multiplet are such that the 
sidelobes never modify the total amplitude. At orbital phases $\pi /2$ and $3 \pi 
/2$, when the stars have zero radial velocity, the sidelobes are in phase with 
each other, but in quadrature, $\pm \pi/2$ radians out of phase with the 
central peak; at orbital phases 0 and $\pi$, when the star reaches maximum or 
minimum radial velocity, the sidelobes are $\pi$ radians out of phase with each 
other and cancel. Thus only frequency modulation occurs with no amplitude 
modulation, as we expect from our initial conditions. It is these phase 
relationships that distinguish frequency modulation multiplets from amplitude 
modulation multiplets, where all members of the multiplet are in phase at the time 
of amplitude maximum.

\subsubsection{Derivation of the binary parameters}
\label{sec:2.4.2}

From the frequency spectrum we can derive the value of $\alpha$, which is, as seen 
in equation\,(\ref{eq:10}), 
\begin{eqnarray}
	\alpha = 
	\left( {{2\pi G}\over{c}} \right)^{1/3}
	{{P_{\rm orb}^{2/3}}\over{P_{\rm osc}}}
	{{m_2\sin i}\over{(m_1+m_2)^{2/3}}} .
\label{eq:21}
\end{eqnarray}
Since $P_{\rm osc}$ is observationally known and $P_{\rm orb}$ is also determined 
from the spacing of the multiplet, the mass function, which is usually derived 
from spectroscopic measurements of radial velocity in a binary system, is 
eventually derived from photometric observations through $\alpha$:
\begin{eqnarray}
	f(m_1,m_2,\sin i) 
	&\equiv&
	{{m_2^3 \sin^3 i}\over{(m_1 + m_2)^2}}
	\nonumber \\
	&=&
	\alpha^3 
	\frac{P_{\rm osc}^{3}}{P_{\rm orb}^{2}}
 	\frac{c^3}{2 \pi G}.
\label{eq:22}
\end{eqnarray}

The distance between the star 1 and the centre of gravity is also deduced from 
$\alpha$:
\begin{eqnarray}
	a_1\sin i ={{P_{\rm osc}}\over{2\pi}} \alpha c,
\label{eq:23}
\end{eqnarray}
as is the radial velocity:
\begin{eqnarray}
	v_{\rm rad, 1} (t) = {{P_{\rm osc}}\over{P_{\rm orb}}} \alpha c \cos\Omega t.
\label{eq:24}
\end{eqnarray}

\subsubsection{The case of $\alpha \ll 1$}
\label{sec:2.4.3}

Most binary stars have $\alpha \ll 1$. In this case, 
$J_1(\alpha) \simeq \alpha/2$, and the value of $\alpha$ is derived to be
\begin{eqnarray}
	\alpha =
	{{A_{+1}+A_{-1}}\over{A_0}}.
\label{eq:25}
\end{eqnarray}
Then, the mass function, the distance between the star and the centre of gravity, 
and the radial velocity are derived to be
\begin{eqnarray}
	f(m_1,m_2,\sin i) 
	=
	\left(\frac{A_{+1}+A_{-1}}{A_0}\right)^3 
	\frac{P_{\rm osc}^{3}}{P_{\rm orb}^{2}}
 	\frac{c^3}{2 \pi G},
\label{eq:26}
\end{eqnarray}
\begin{eqnarray}
	a_1\sin i = {{P_{\rm osc}}\over{2\pi}} {{A_{+1}+A_{-1}}\over{A_0}} c,
\label{eq:27}
\end{eqnarray}
and
\begin{eqnarray}
	v_{\rm rad, 1}(t) = {{P_{\rm osc}}\over{P_{\rm orb}}} 
	{{A_{+1}+A_{-1}}\over{A_0}} c \cos\Omega t.
\label{eq:28}
\end{eqnarray}

\subsection{An example with artificial data for the case of a circular orbit}
\label{sec:2.5}

\begin{figure}
\centering
\includegraphics[width=\linewidth,angle=0]{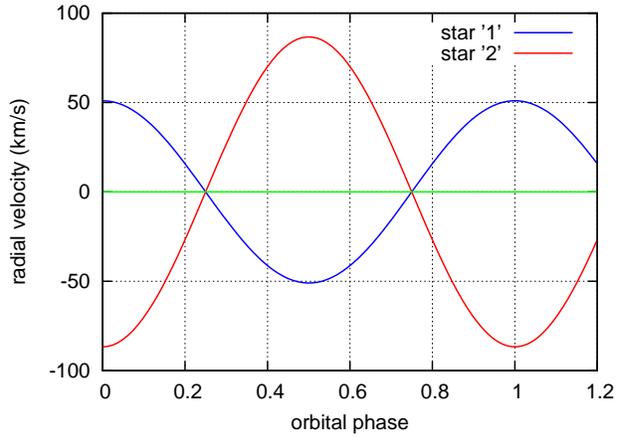}
\caption{An artificial radial velocity curve for a binary system with a 
1.7-M$_{\odot}$ $\delta$\,Sct star in a 10-d circular orbit with a 1-M$_{\odot}$ 
companion; $i = 90^\circ$. The blue curve is for the primary component and the red 
curve the secondary.}
\label{fig:5}
\end{figure}

We illustrate the results derived in the previous subsection with artificial data 
generated for the following parameters: $m_1 = 1.7$\,M$_{\odot}$ and $\nu_{\rm 
osc} \equiv 1/P_{\rm osc} = 20$\,d$^{-1}$ -- typical of a late-A $\delta$\,Sct 
star; $e=0$, of course; $m_2 = 1$\,M$_{\odot}$; $P_{\rm orb} = 10$\,d; and $i = 
90^\circ$. In this case our parameter $\alpha = 3.39 \times 10^{-2}$ (see 
equation\,(\ref{eq:10}) and Fig.\,\ref{fig:2}). Fig.\,\ref{fig:5} shows the radial 
velocity curve for this system where our convention is the star 1 -- the pulsating 
A star -- is at maximum velocity of recession at phase zero. These represent a 
typical eclipsing binary $\delta$\,Sct Am star. (The reason we say Am star in this 
case is that most A star binary systems with $P_{\rm orb} = 10$\,d have 
synchronous rotation, leading to the slow rotation that is a prerequisite for 
atomic diffusion in metallic-lined A stars.)

\begin{figure}
\centering
\includegraphics[width=\linewidth,angle=0]{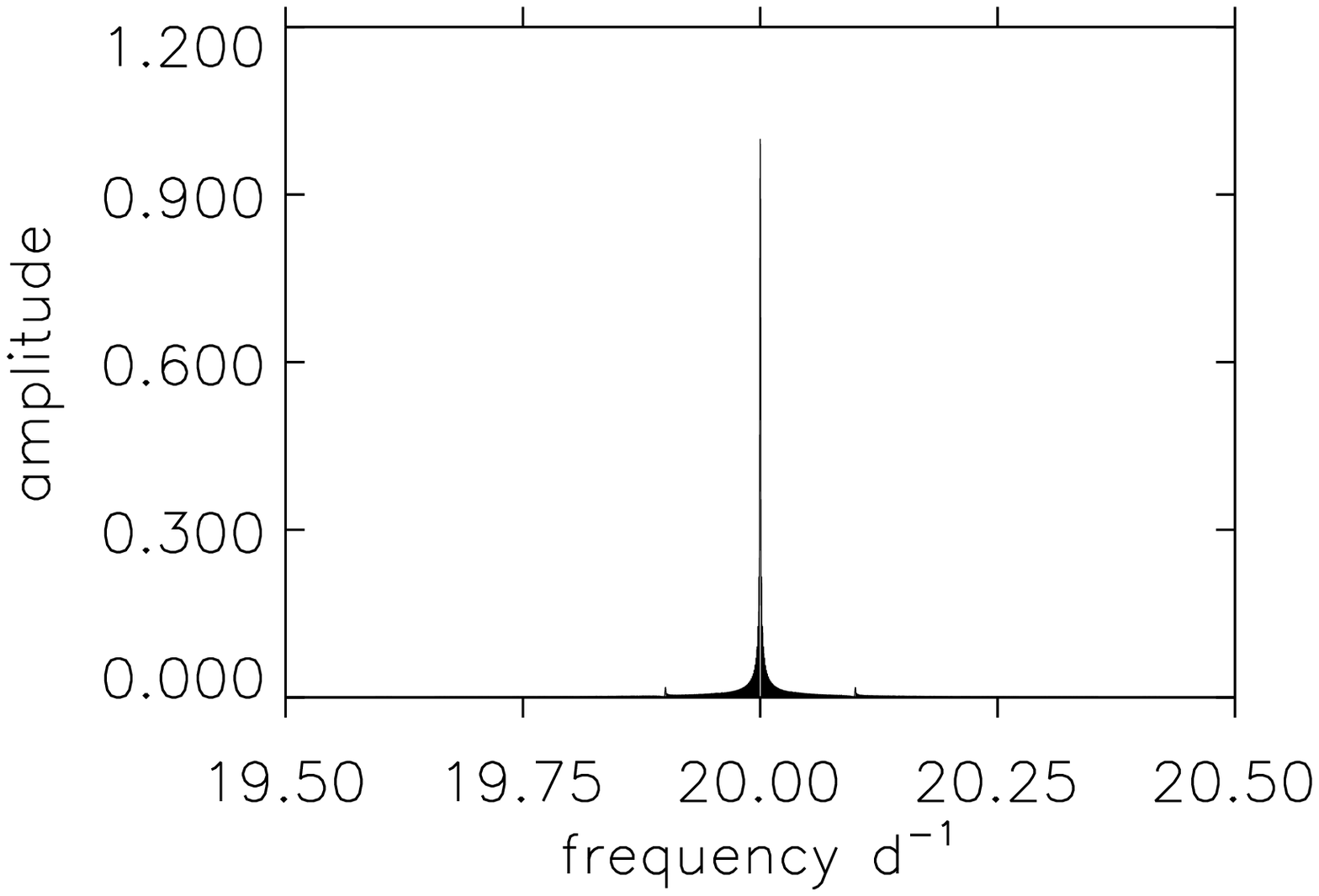}
\includegraphics[width=\linewidth,angle=0]{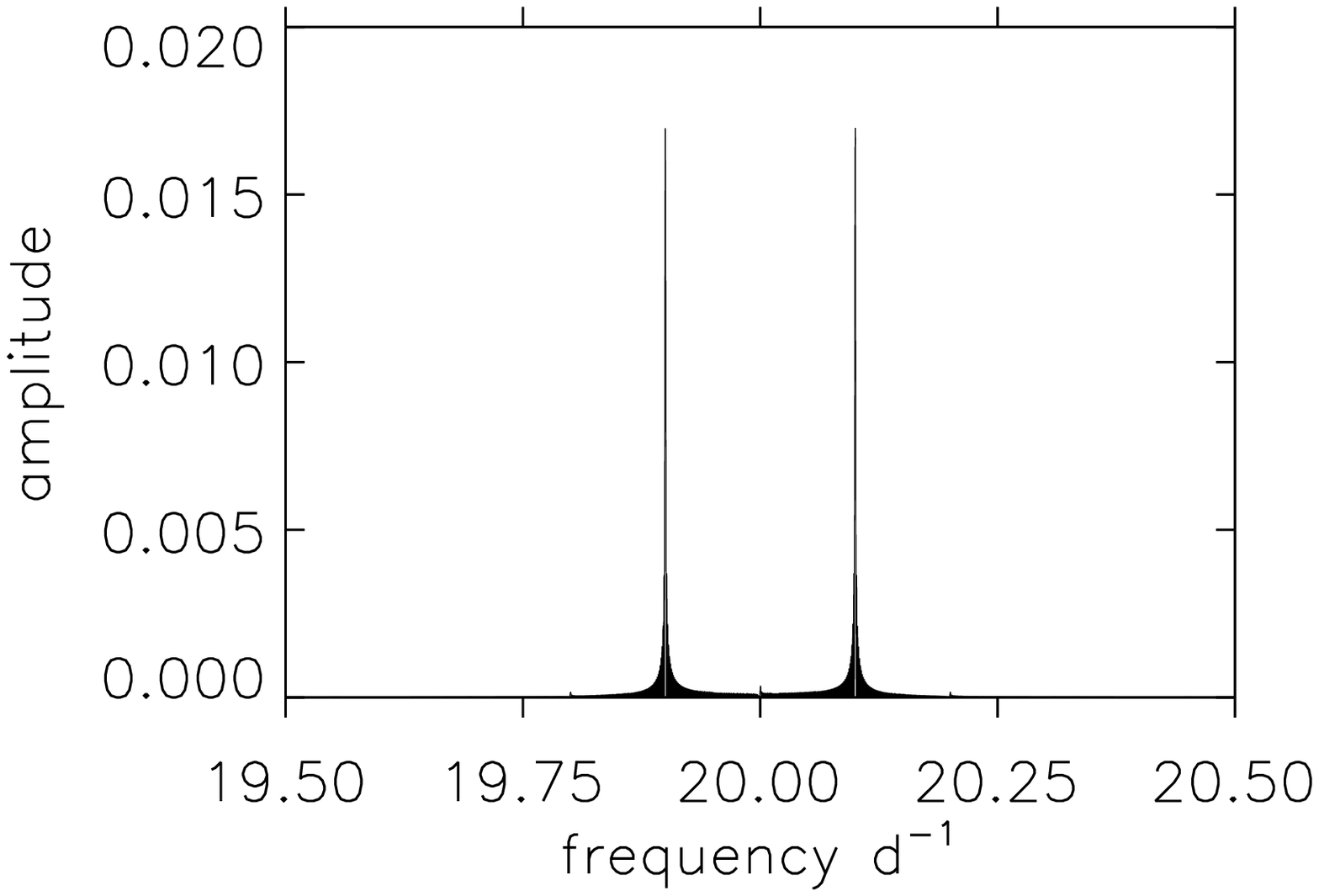}
\caption{Top panel: An amplitude spectrum of the artificial data around the chosen 
pulsation frequency of 20\,d$^{-1}$. The amplitude has been chosen to be in 
intensity units. The first sidelobes at $\pm \nu_{\rm orb} \,(\equiv\pm 1/P_{\rm 
orb})$ are barely visible. Bottom panel: The amplitude spectrum after prewhitening 
by the central peak of the multiplet. The two first sidelobes are evident with 
amplitudes of 0.017 of the amplitude of the central peak. The second sidelobes are 
insignificant. }
\label{fig:6}
\end{figure}

We have generated an artificial light curve with no noise using 10 points per 
pulsation cycle and a time span of 10 orbital periods (100\,d) in a hare-and-hound 
test to see how well the input binary parameters are reproduced from the 
light curve. The top panel of Fig.\,\ref{fig:6} shows an amplitude spectrum of the 
generated light curve around the chosen pulsation frequency, 20\,d$^{-1}$. The 
first sidelobes at 19.9\,d$^{-1}$ and 20.1\,d$^{-1}$ are barely visible in this 
panel, but are clearly seen after prewhitening the central peak as shown in the 
bottom panel of Fig.\,\ref{fig:6}. They are separated from that central peak by 
0.1\,d$^{-1}$ and to have relative amplitudes of 0.017. The orbital period, 10\,d, 
is well determined from the spacing. From the amplitude ratio, the value of 
$\alpha$ is also reasonably well reproduced.
 
\begin{table}
\centering
\caption[]{Least squares fit of the frequency triplet to the artificial data at 
the orbital phase of eclipse for the orbital parameters given in 
Section\,\ref{sec:2.5}. The two sidelobes are in phase with each other, but 
$90^\circ$ out of phase with the central peak, as expected. The frequencies are 
split by the orbital frequency, given $P_{\rm orb} = 10$\,d. These are artificial 
data with no noise added, hence there are no errors.}
\small
\begin{tabular}{ccc}
\hline
\multicolumn{1}{c}{frequency} & \multicolumn{1}{c}{amplitude} &   
\multicolumn{1}{c}{phase} \\
\multicolumn{1}{c}{d$^{-1}$} &  &   
\multicolumn{1}{c}{radians}\\
\hline
  19.9     &     0.017    &     1.57 \\
  20.0     &     1.000    &     0.00 \\
  20.1     &     0.017    &     1.57 \\
\hline
\end{tabular}
\label{table:1}
\end{table}

From the amplitude ratio in Table\,1 and equation (\ref{eq:22}) we derive $f(m_1, 
m_2, \sin i) = 0.137\,{\rm M}_{\odot}$, as expected for the input parameters of $i 
= 90^\circ$, $m_1 = 1.7\,{\rm M}_{\odot}$ and $m_2 = 1.0\,{\rm M}_{\odot}$. Hence 
we have shown that the mass function can be derived entirely from the photometric 
light curve. The expected radial velocity of star 1 is derived from equation 
(\ref{eq:24}), from which its amplitude is determined to be 51\,km\,s$^{-1}$. 
Hence the radial velocity curve, shown with the blue curve in Fig.\,\ref{fig:5}, 
has also been well reproduced only from the photometric light curve. 

\subsection{An actual example for the case of a circular orbit: the hierarchical 
multiple system KIC\,4150611 = HD\,181469}
\label{sec:2.6}

Let us now look at an actual example. The {\it Kepler} mission is observing about 
150\,000 stars over its 115 square degree field-of-view for time spans of one 
month to years. In the data for {\it Kepler} `quarters' (1/4 of its 372-d 
heliocentric orbit) Q1 to Q9 we can see a hierarchical multiple star system of 
complexity and interest, KIC\,4150611. This system is composed of an eccentric 
eclipsing binary pair of G stars in an 8.6-d orbit that are a common proper motion 
pair with a $\delta$\,Sct A star in a circular orbit about a pair of K stars with 
an orbital period of $94.1 \pm 0.1$\,d; the K star binary itself has an orbital 
period of about 1.5\,d. These five stars show a remarkable set of eclipses, 
successfully modelled by the {\it Kepler} eclipsing binary star working group 
(Pr\v sa et al., in preparation). Here we show that we can derive the mass 
function for the A star -- K-binary system from the light curve alone by using the 
$\delta$\,Sct pulsations as a clock. This is an important advance for a system 
such as this. Measuring radial velocities from spectra requires a great effort at 
the telescope to obtain the spectra. Relatively low accuracy then results from 
 spectral disentangling of the A star 
from the other components of the system, and as a 
consequence of the rotational velocity of the A star of $v \sin i \sim 
100$\,km\,s$^{-1}$. Using our technique the A star is the only pulsating star in 
the system, hence the photometric radial velocities come naturally from just the A 
star primary and we are unaffected by rotational broadening or spectral 
disentangling.  

\begin{figure}
\centering
\includegraphics[width=\linewidth,angle=0]{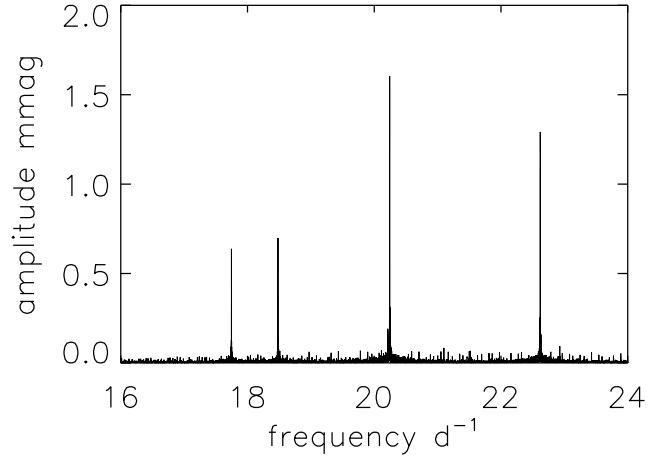}
\caption{An amplitude spectrum for the Q1 to Q9 {\it Kepler} long cadence data in 
the $\delta$\,Sct frequency range of the A-star component of the multiple system 
KIC\,4150611. The four peaks are from independent, low overtone modes.  }
\label{fig:7}
\end{figure}

The data we use are Q1 to Q9 long cadence {\it Kepler} data with integration times 
of 29.4\,min covering a time span of 774\,d. The Nyquist frequency for these data 
is about 24.5\,d$^{-1}$. We also have short cadence {\it Kepler} data for this 
star (integration times of 58.9\,s) that show the $\delta$\,Sct pulsations have 
frequencies less than the long cadence Nyquist frequency. We have masked the 
eclipses in the light curve and run a high-pass filter, leaving only the 
$\delta$\,Sct pulsation frequencies. Fig.\,\ref{fig:7} shows an amplitude spectrum 
for these data where four peaks stand out. Broad-band photometric data from the 
{\it Kepler} Input Catalogue photometry suggests $T_{\rm  eff} \approx 6600$\,K 
and $\log g = 4$ (in cgs units) for this star, but this temperature is likely to 
be underestimated because of the light of the cooler companion stars. We therefore 
estimate that the A star has a spectral type around the cool border of the 
$\delta$\,Sct instability strip, $T_{\rm  eff} \sim 7400$\,K.

\begin{table}
\centering
\caption[]{A non-linear least-squares fit of the four highest amplitude 
frequencies seen in Fig.\,\ref{fig:7} to the Q1 to Q9 {\it Kepler} data for 
KIC\,4150611. The range of frequencies suggests low overtone modes. The zero point 
in time for the phase is BJD\,2455311.758.}
\small
\begin{tabular}{ccc}
\hline
\multicolumn{1}{c}{frequency} & \multicolumn{1}{c}{amplitude} &   
\multicolumn{1}{c}{phase} \\
\multicolumn{1}{c}{d$^{-1}$} & \multicolumn{1}{c}{mmag} &   
\multicolumn{1}{c}{radians}\\
\hline
$17.746558 \pm 0.000008$ & $0.640  \pm 0.007$ & $\phantom{-}1.894  \pm 0.011$  \\
$18.480519 \pm 0.000007$ & $0.699  \pm 0.007$ & $-0.029  \pm 0.010$ \\
$20.243260  \pm 0.000003$ & $1.610 \pm 0.007$ & $\phantom{-}2.360 \pm 0.004$ \\
$22.619577 \pm 0.000004$ & $1.294  \pm 0.007$ & $\phantom{-}0.150 \pm 0.005$  \\
\hline
\end{tabular}
\label{table:2}
\end{table}

\begin{table*}
\centering
\caption[]{A least squares fit of the frequency triplets for the four high 
amplitude modes to the Q1 to Q9 {\it Kepler} data for KIC\,4150611. The 
frequencies of the multiplets are separated by the orbital frequency. The zero 
point for the phases has been chosen to be a time of transit of the A star by its 
companions, $t_0 = {\rm BJD}\,2455311.758$. Column 4 shows that the phases of the 
sidelobes are equal within the errors at this time and column 5 shows that they 
are $\pi/2 = 1.57$ radians out of phase with the central peak. Column 6 shows the 
amplitude ratios of the sidelobes to the central peaks. Since they are small 
compared to unity, they are equal to $\alpha$ to great accuracy. The ratios of 
$\alpha$ to the frequency are the same for all modes, as shown in column 7. The 
phase relations and amplitude ratios are as expected from our theory.}
\small
\begin{tabular}{ccccccc}
\hline
\multicolumn{1}{c}{frequency} & \multicolumn{1}{c}{amplitude} &   
\multicolumn{1}{c}{phase} & \multicolumn{1}{c}{$\phi_{+1}-\phi_{-1}$} & 
\multicolumn{1}{c}{${\langle\phi_{+1}-\phi_{-1}\rangle}-{\phi_0}$} & 
\multicolumn{1}{c}{$\frac{A_{+1}+A_{-1}}{A_0}$} & 
$\alpha/\nu_{\rm osc}$ 
\\
\multicolumn{1}{c}{d$^{-1}$} & \multicolumn{1}{c}{mmag} &   
\multicolumn{1}{c}{radians} & \multicolumn{1}{c}{radians} & 
\multicolumn{1}{c}{radians} & & 
$\times 10^{-3}$\,d 
\\
\hline
17.735937  &  $0.049  \pm 0.007$  &  $\phantom{-}0.299  \pm 0.139$ &  &  &  &  \\
17.746558  &  $0.640  \pm 0.007$  &  $\phantom{-}1.892  \pm 0.011$ & 
$-0.04 \pm 0.24$ & $-1.61 \pm 0.12$  & $0.133 \pm 0.016$ & $7.49 \pm 0.90$
\\
\vspace{2mm}17.757179  &  $0.036  \pm 0.007$  &  $\phantom{-}0.258  \pm 0.191$ &  
&  &  &  \\
18.469898  &  $0.050  \pm 0.007$  &  $-1.654  \pm 0.136$           &  &  &  &\\
18.480519  &  $0.699  \pm 0.007$  &  $-0.031  \pm 0.010$           & 
$\phantom{-}0.04 \pm 0.20$ &  $-1.60 \pm 0.10$  & $0.139 \pm 0.014$ & $7.52 \pm 
0.75$ 
\\
\vspace{2mm}18.491140  &  $0.047  \pm 0.007$  &  $-1.617  \pm 0.147$           &  
&  &  &  \\
20.232639  &  $0.117  \pm 0.007$  &  $\phantom{-}0.812  \pm 0.058$ &  &  &  &  \\
20.243260  &  $1.610  \pm 0.007$  &  $\phantom{-}2.358  \pm 0.004$ & $-0.11 
\pm0.08$ &  $-1.60 \pm 0.40$  & $0.148 \pm 0.006$ & $7.31 \pm 0.30$ 
\\
\vspace{2mm}20.253881  &  $0.122  \pm 0.007$  &  $\phantom{-}0.699  \pm 0.056$ &  
&  &  &  \\
22.608956  &  $0.106  \pm 0.007$  &  $-1.441  \pm 0.065$           &  &  &  &  \\
22.619577  &  $1.294  \pm 0.007$  &  $\phantom{-}0.147  \pm 0.005$ & $-0.04 \pm 
0.09$  &  $-1.61 \pm 0.47$  & $0.162 \pm 0.008$ & $7.16 \pm 0.35$
\\
22.630198  &  $0.104  \pm 0.007$  &  $-1.485  \pm 0.066$           &  &  &  \\
\hline
\end{tabular}
\label{table:3}
\end{table*}

Table\,\ref{table:2} shows the frequencies, amplitudes and phases for the four 
peaks seen in Fig.\,\ref{fig:7}. For a first estimate of mode identification, it 
is useful to look at the $Q$ value for each of the four frequencies. This is 
defined to be
\begin{equation}
	Q = {P_{\rm osc}}\sqrt{\frac{\overline{\rho}}{\overline{\rho_{\odot}}}}
\label{eq:29}
\end{equation}
where $P_{\rm osc}$ is the pulsation period and $\overline\rho$ is the 
mean density; $Q$ is known as the `pulsation constant'. 
Equation\,(\ref{eq:29}) can be rewritten as
\begin{equation}
\log Q = -6.454 +\log P_{\rm osc} +\frac{1}{2}\log g +\frac{1}{10}M_{\rm bol} + 
\log T_{\rm eff}, 
\label{eq:30}
\end{equation}
where $P_{\rm osc}$ is given in d, $\log g$ uses cgs units and $T_{\rm  eff}$ is 
in Kelvin. Using $T_{\rm  eff} = 7400$\,K and $\log g = 4.0$, and estimating the 
bolometric magnitude to be about 2 gives $Q$ values in the range 0.019 to 0.023. 
Standard values for $\delta$\,Sct models are $Q = 0.033$ for the fundamental mode 
and $Q = 0.025$ for the first overtone, thus suggesting that the four modes so far 
examined have radial overtones higher than that. There are additional pulsation 
mode frequencies of low amplitude that are not seen at the scale of this figure. 
Those will be examined in detail in a future study.

\begin{figure}
\centering
\includegraphics[width=\linewidth,angle=0]{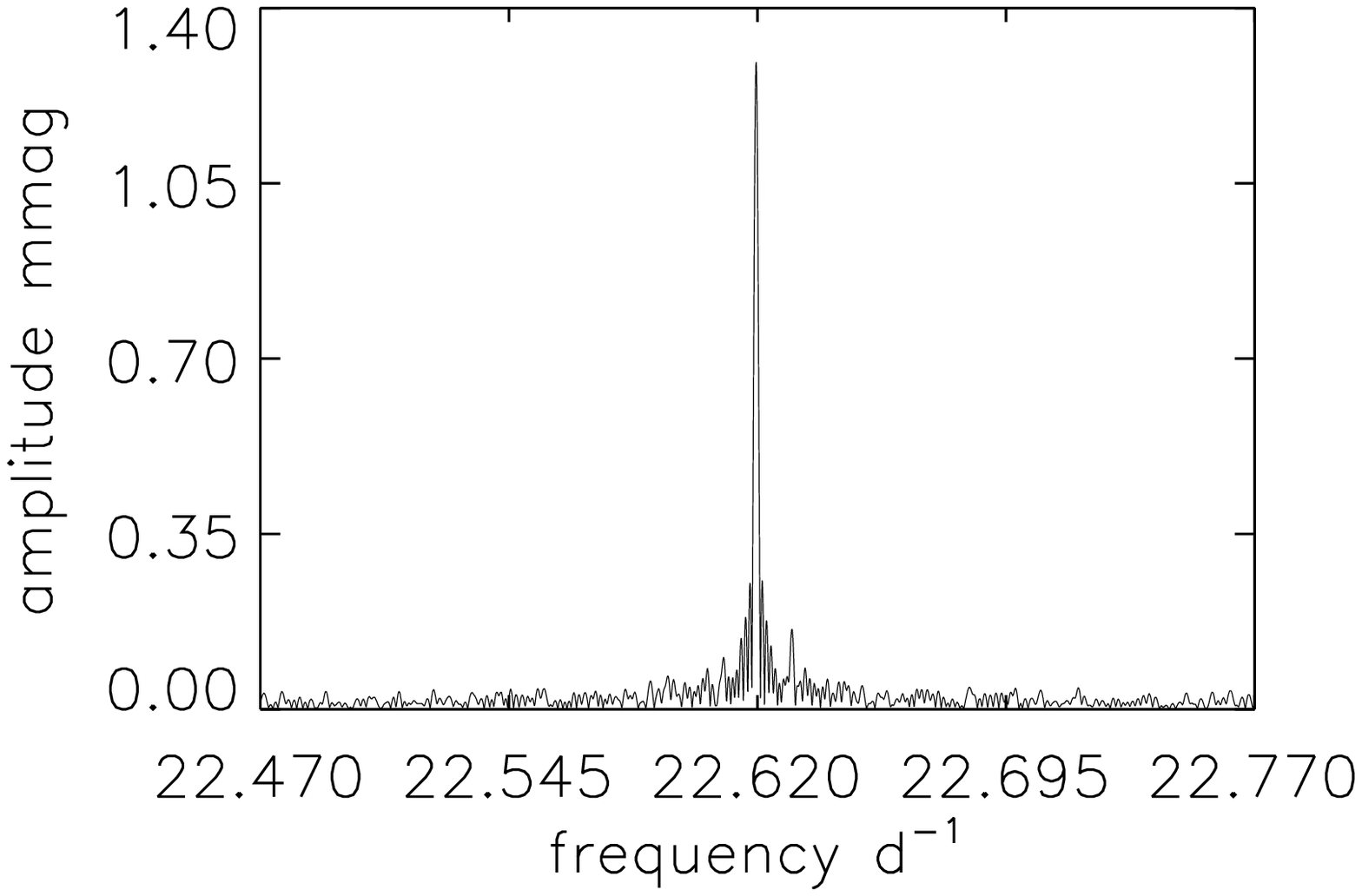}
\includegraphics[width=\linewidth,angle=0]{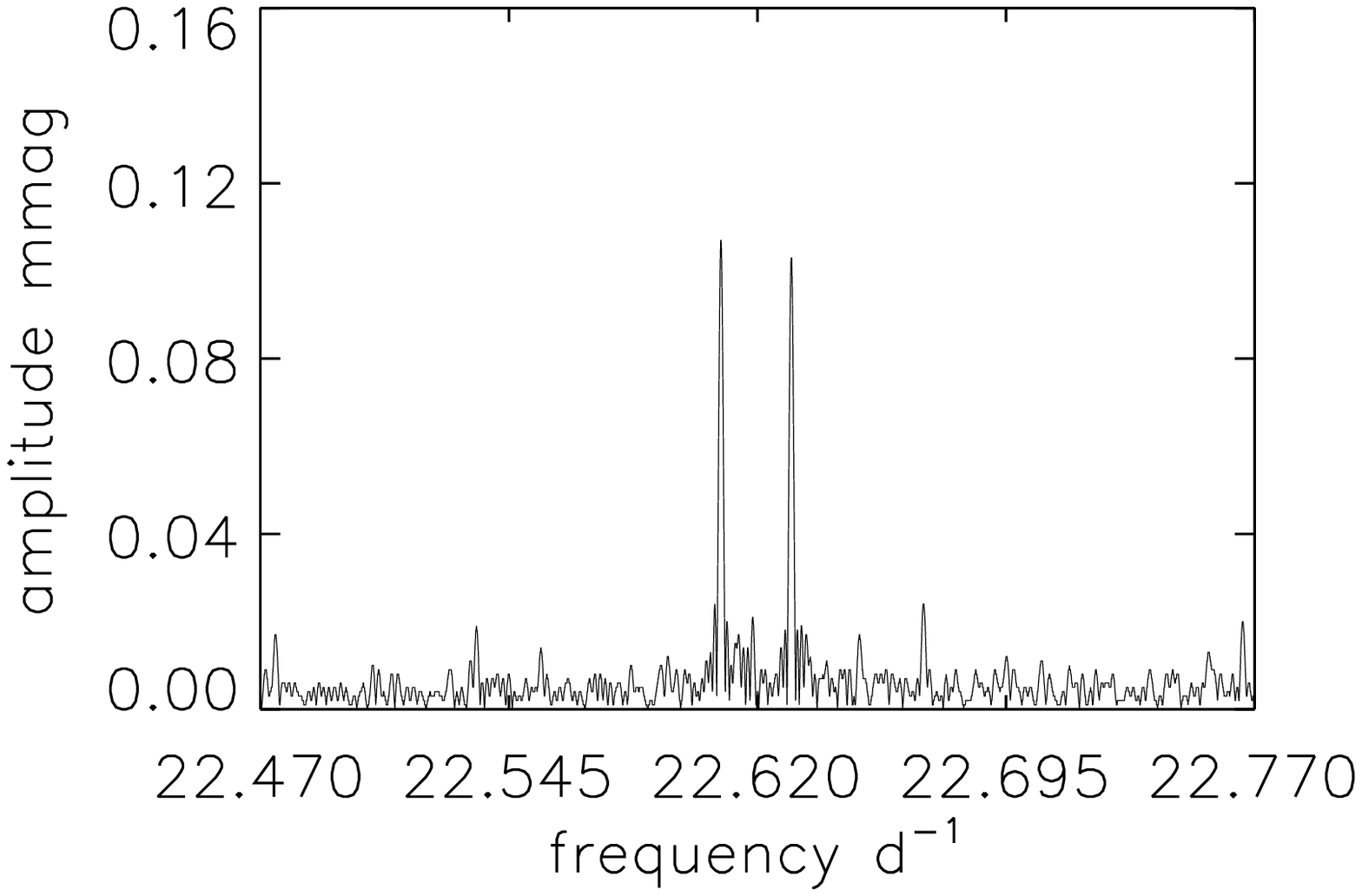}
\caption{Top panel: An amplitude spectrum for the Q1 to Q9 KIC\,4150611 data 
centred on the peak at 22.619577\,d$^{-1}$. Bottom panel: An amplitude spectrum 
after the central peak has been prewhitened, showing the two sidelobes split from 
the central peak by exactly the orbital frequency.}
\label{fig:8}
\end{figure}

What we wish to examine here are the sidelobes to these four highest amplitude 
peaks. Each of these shows a frequency triplet split by the orbital frequency. The 
highest amplitude peak at 20.243260\,d$^{-1}$ also has another pulsation mode 
frequency nearby, so we illustrate the triplets with the simpler example of the 
second-highest peak at 22.619577\,d$^{-1}$ as shown in Fig.\,\ref{fig:8}. 

There is clearly an equally spaced triplet for this frequency, and this is the 
case for all four mode frequencies. Table\,\ref{table:3} shows a least-squares fit 
of the frequency triplets for the four modes. After fitting the data by nonlinear 
least-squares with the four triplets and showing that the frequency spacing is the 
same within the errors in all cases, we forced each triplet to have exactly equal 
spacing with a separation of the average orbital frequency determined from all 
four triplets. To examine the phase relationship of the triplet components, it is 
important to have exactly equal splitting because of the many cycles back to the 
time zero point. From the separation of the triplet components, we derive the 
orbital period of the star is $94.09 \pm 0.11$\,d.

It can be seen in Table\,\ref{table:3} that the data are an excellent fit to our 
theory. The zero point in time has been selected to be a time of transit in this 
eclipsing system as seen in the light curve. The expectation is that the sidelobes 
should be in phase at this time and exactly $\pi/2$ radians out of phase with the 
central peak. That is the case for all four triplets. Since the amplitude ratios 
of the sidelobes to the central peaks of each triplet are small compared to unity, 
they are regarded as $\alpha$ with great accuracy. It is expected that the ratio 
of $\alpha$ to the frequency is the same for all triplets, since this is directly 
proportional to the mass function, as in equation\,(\ref{eq:22}). 
Table\,\ref{table:3} shows that this is the case. By using the values obtained for 
$\alpha$, $P_{\rm orb}$ and $P_{\rm osc}$, we deduce the projected radius of the 
orbit, $a_1\sin i$, the radial velocity, $v_{\rm rad, 1}$ and the mass function. 
Table\,\ref{table:4} gives these values derived from each triplet. The consistency 
of the values for the four independent pulsation frequencies is excellent.

\begin{table}
\centering
\caption[]{The values of $a_1\sin i$, the amplitude of the radial velocity and the 
mass function derived from each triplet. The consistency of the values derived for 
the four triplets is substantially better than the formal errors might lead us to 
expect, suggesting that those formal errors may be overestimated.}
\begin{tabular}{cccc}
\hline
frequency & $a_1\sin i$ & RV amplitude & $f(m_1,m_2,\sin i)$ \\
d$^{-1}$ & au & km\,s$^{-1}$ & M$_\odot$ \\
\hline
17.746558 & $1.30 \pm 0.16$ & $23.9 \pm 2.9$ & $0.132 \pm 0.116$\\
18.480519 & $1.30 \pm 0.13$ & $23.9 \pm 2.4$ & $0.133 \pm 0.098$\\
20.243260 & $1.27 \pm 0.05$ & $23.3 \pm 1.0$ & $0.124 \pm 0.035$\\
22.619577 & $1.24 \pm 0.06$ & $22.8 \pm 1.1$ & $0.116 \pm 0.034$\\
\hline
\end{tabular}
\label{table:4}
\end{table}

Because of the better signal-to-noise ratio for the two highest amplitude 
frequencies, we derive from them a best estimate of the mass function of $f(m_1, 
m_2, \sin i) = 0.120 \pm 0.024$\,M$_\odot$. If instead of propagating errors we 
take the average and standard deviation of the four values of the mass function 
from Table\,\ref{table:4}, this gives a value of $f(m_1, m_2, \sin i) = 0.126 \pm 
0.008$\,M$_\odot$. Assuming a mass of $M \sim 1.7$\,M$_{\odot}$ for the A star 
then gives a total mass for the two K stars of the 1.5-d binary companion (see 
Pr\v sa et al., in preparation) less than 1\,M$_{\odot}$. We have determined the 
mass function in KIC\,4150611 entirely from the photometric light curve by using 
the $\delta$\,Sct pulsations as clocks and extracting the information from the 
light time effect by means of the Fourier transform. This is a significant 
improvement to what can be done for this star with spectroscopic radial 
velocities. 

\begin{figure}
\centering
\includegraphics[width=\linewidth,angle=0]{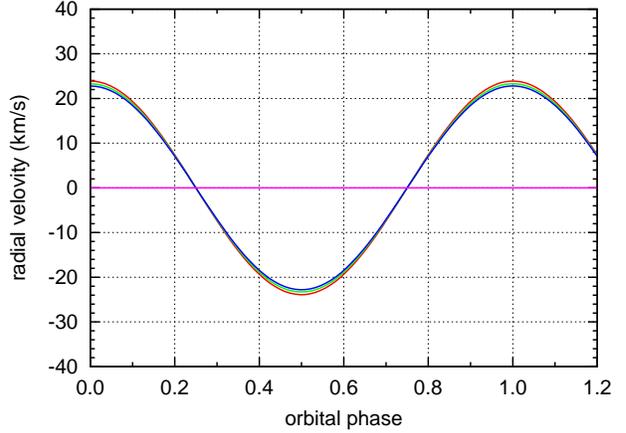}
\caption{The radial velocity curves of KIC\,4150611 derived entirely from the 
photometric light curve by using the $\delta$\,Sct pulsations as clocks and 
extracting the information from the light time effect by means of the Fourier 
transform. Note that different curves correspond to different multiplets, but the 
differences are small.}
\label{fig:9}
\end{figure}

Fig.\,\ref{fig:9} shows the radial velocity curves of KIC\,4150611 derived from 
the four sets of multiplets. Note that the differences are small. This result will 
be tested by comparison with the radial velocities obtained with spectroscopic 
observations by Pr\v sa et al. (in preparation). An advantage of the present 
analysis is that photometric observations by {\it Kepler} have covered a long time 
span with few interruptions. The duty cycle is far superior to what is currently 
possible with spectroscopic observations.

So far we have assumed that the A-star-K-binary system of KIC\,4150611 has a 
circular orbit. How do we justify this assumption? To find out, we now return to 
the theory for cases more complex than a circular orbit. 

%%%%%%%%%%%%%%%%%%%%%%%%%%%%%%%%%%%%%%
\section{The theory for the general case of eccentric orbits}
\label{sec:3}

Now let us consider a more realistic case: elliptical orbital motion. The radial 
velocity curve deviates from a pure sinusoidal curve with a single period. Instead 
of a simple sinusoid, it is expressed with a Fourier series of the harmonics of 
the orbital period. 

\subsection{Radial velocity along the line of sight}
\label{sec:3.1}

\begin{figure}
\centering
\includegraphics[width=0.6\linewidth,angle=0]{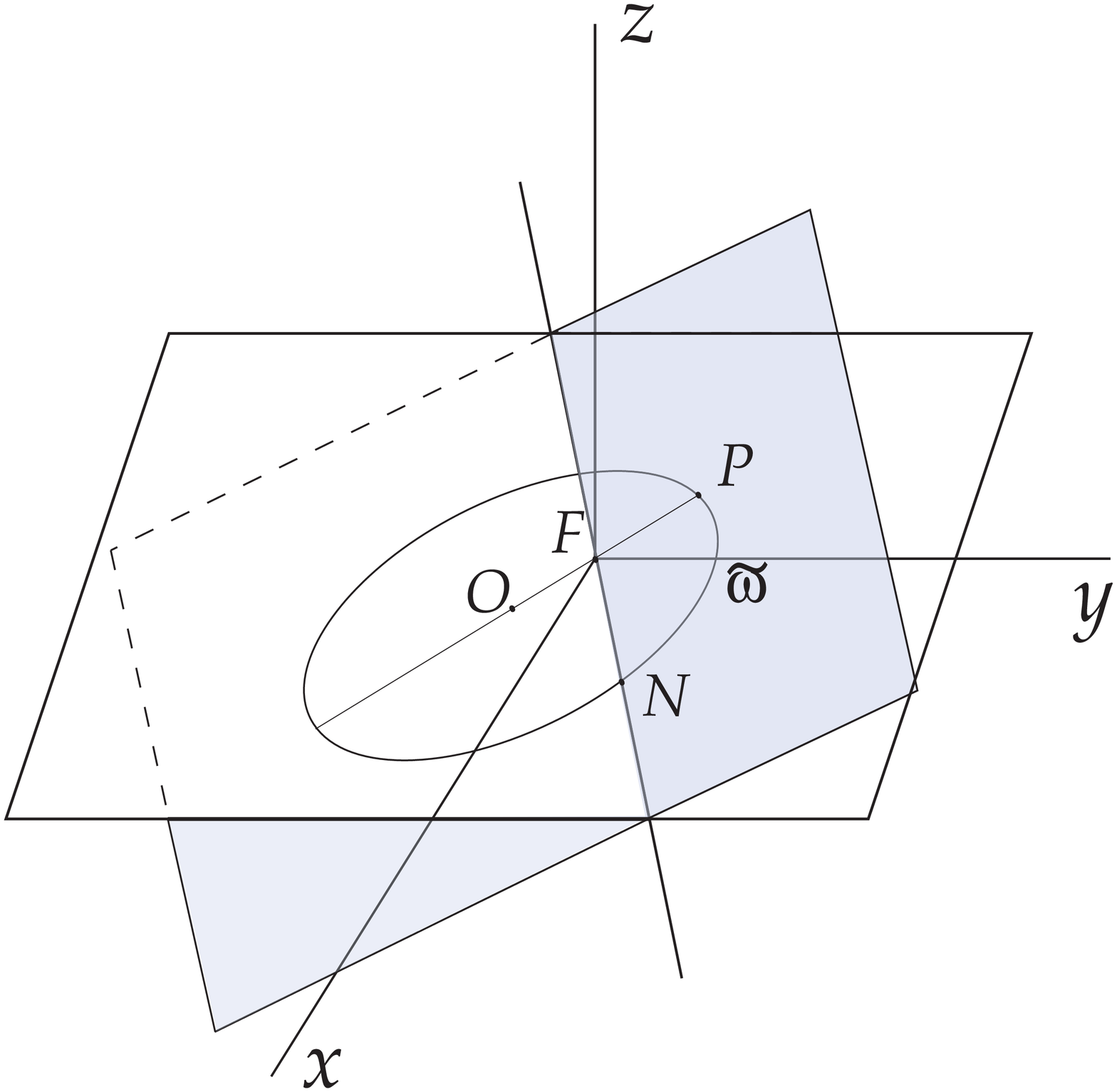}
\includegraphics[width=0.6\linewidth,angle=0]{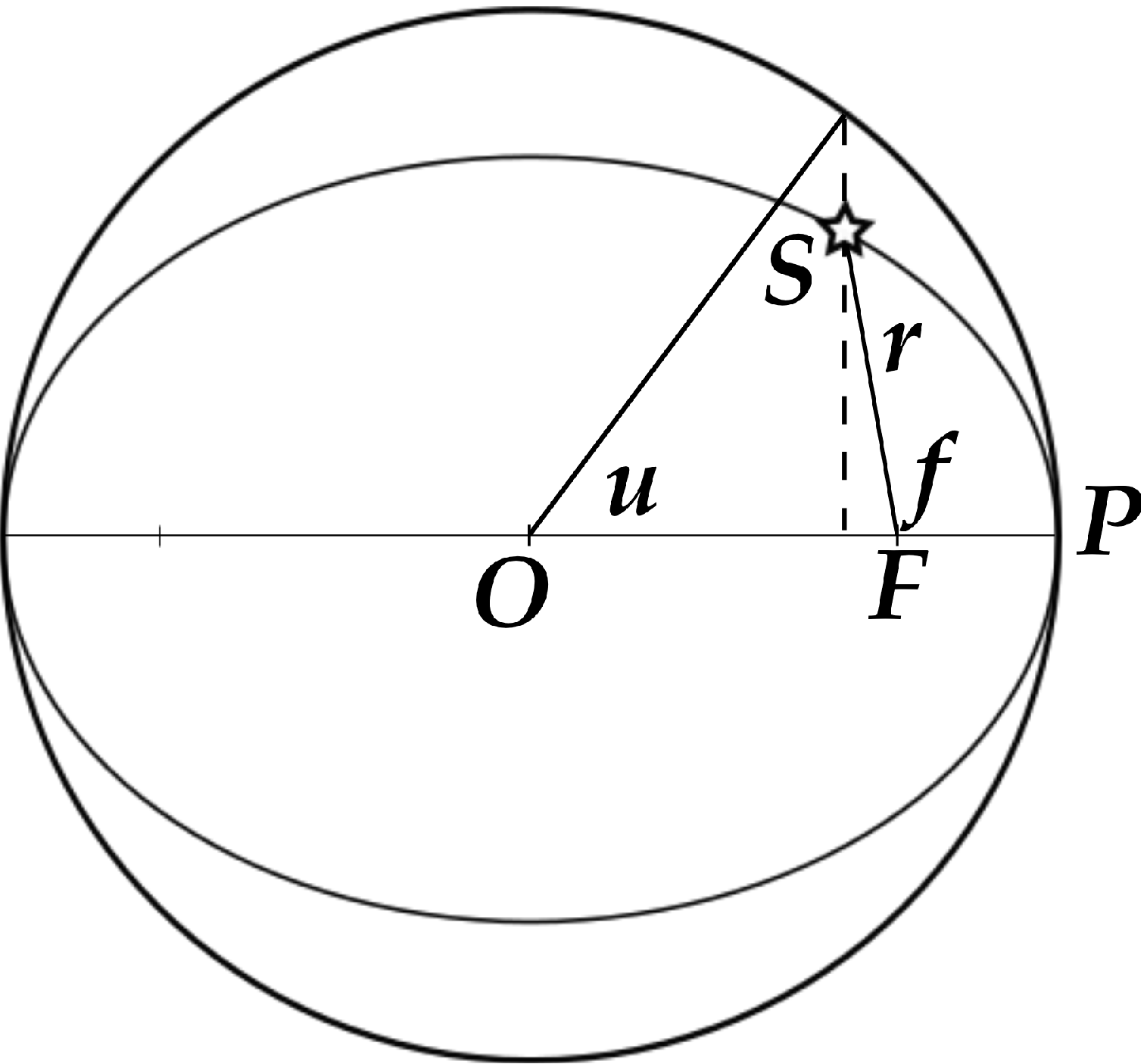}
\caption{Geometrical configuration. Top: The $xy$-plane is tangent to the 
celestial sphere, and the $z$-axis is the line of sight toward us. 
The origin `F' is the 
focus of the orbital ellipse; that is, the centre of gravity of the binary system. 
The orbital plane is inclined to the $xy$-plane by the angle $i$. The periapsis of 
the ellipsoidal orbit is `P', and the ascending node is `N'. The angle NFP is 
$\varpi$. Bottom: Schematic top view 
(i.e., along the normal to the orbital plane) of the 
orbital plane. The star is located, at this moment, at `S' on the orbital ellipse, 
for which the focus is `F'. The semi-major axis is $a_1$ and the eccentricity is 
$e$. Then $\overline{{\rm OF}}$ is $a_1 e$. The distance between the focus, F, and 
the star, S, is $r$. The angle PFS is `the true anomaly', $f$. `The eccentric 
anomaly', $u$, is defined through the circumscribed circle that is concentric 
with the orbital ellipse.}
\label{fig:10}
\end{figure}

Let the $xy$-plane be tangent to the celestial sphere, and let the $z$-axis, being 
perpendicular to the $xy$-plane and passing through the centre of gravity of the 
binary, be along the line of sight toward us. 
The orbital plane of the binary motion is 
assumed to be inclined to the $xy$-plane by the angle $i$. The orbit is an 
ellipse. We write the semi-major axis and the eccentricity of the orbit as $a_1$ and 
$e$, respectively. Let $\varpi$ be the angle between the ascending node, which is 
an intersection of the orbit and the $xy$-plane, and the periapsis. Also let $f$ 
be the angle between the periapsis and the star at the moment, that is the `true 
anomaly', and let $r$ be the distance between the centre of gravity and the star 
(see Fig.\,\ref{fig:10}). 

Then, the $z$-coordinate of the position of the star is written as
\begin{eqnarray}
	z = r\sin (f+\varpi) \sin i .
\label{eq:31}
\end{eqnarray}
The radial velocity along the line of sight, 
$v_{\rm rad, 1} = -{\rm d}z/{\rm d}t$, is then
\begin{eqnarray}
	v_{\rm rad, 1} = 
	- \left[ {{{\rm d}r}\over{{\rm d}t}} \sin (f+\varpi) 
	+ r {{{\rm d}f}\over{{\rm d}t}} \cos (f+\varpi) \right]\sin i.
\label{eq:32}
\end{eqnarray}
It should be noted here again that 
the sign of $v_{\rm rad, 1}$ is defined so that $v_{\rm rad, 1} > 0$ when the star 
is receding from us.
The distance $r$ between the focus `F' and the star `S' is expressed with help of 
a combination of the semi-major axis $a_1$, the eccentricity $e$ and the true 
anomaly $f$:
\begin{eqnarray}
	r = {{a_1 (1-e^2)}\over{1+e\cos f}}.
\label{eq:33}
\end{eqnarray}
From the known laws of motion in an ellipse (see text books; e.g., 
\citealt{brouwerclemence1961}), we have
\begin{eqnarray}
	r{{{\rm d}f}\over{{\rm d}t}} = {{a_1 \Omega (1+e\cos f)}\over{\sqrt{1-e^2}}}
\label{eq:34}
\end{eqnarray}
and
\begin{eqnarray}
	{{{\rm d}r}\over{{\rm d}t}} = {{a_1 \Omega e\sin f}\over{\sqrt{1-e^2}}}.
\label{eq:35}
\end{eqnarray}
Therefore, the radial velocity of the star 1 along the line of sight is 
expressed as
\begin{eqnarray}
	v_{\rm rad, 1}
	&=&
	-{{\Omega a_1\sin i}\over{\sqrt{1-e^2}}} 
	\,\left[\cos (f+\varpi) + e\cos\varpi\right]
\label{eq:36}
	\\
	&=& 
	-(2\pi G{\rm M}_\odot)^{1/3} \left({{m_1}\over{{\rm M}_\odot}}\right)^{1/3} 
	q (1+q)^{-2/3} P_{\rm orb}^{-1/3} \sin i
	\nonumber \\
	& & 
	\times
	{{1}\over{\sqrt{1-e^2}}}
	\,\left(
	\cos f\cos\varpi - \sin f\sin\varpi
	\right.
	\nonumber \\
	& &
	\left.
	 + e\cos\varpi \right).
\label{eq:37}
\end{eqnarray}
In the case of $e=0$, the periapsis is not uniquely defined, nor 
are the angles $\varpi$ and $f$. Instead, 
the angle between the ascending node and the star at the moment, $(f+\varpi)$, 
is well defined.
If we choose $\varpi=\pi$, $f$ means the angle between the descending node and the 
star at the moment.

\subsection{Phase modulation}
\label{sec:3.2}
\subsubsection{General formulae}
\label{sec:3.2.1}
To evaluate phase modulation, we have to integrate the radial velocity with 
respect to time. The time dependence of radial velocity is implicitly expressed by 
the true anomaly $f$, which can be written in terms of `the eccentric anomaly', 
$u$ (see Fig.\,\ref{fig:10}), defined through the circumscribed circle that is 
concentric with the orbital ellipse, as
\begin{eqnarray}
	\cos f = {{\cos u-e}\over{1-e\cos u}}.
\label{eq:38}
\end{eqnarray}
Kepler's equation links the eccentric anomaly $u$ with `the mean anomaly' $l$:
\begin{eqnarray}
	l &\equiv& \Omega (t-t_0)
\nonumber	\\
	&=&
	u-e\sin u,
\label{eq:39}
\end{eqnarray}
where $t_0$ denotes the time of periapsis passage. Various methods of solving 
Kepler's equation to obtain $u$ for a given $l$ have been developed. One of them 
is Fourier expansion. In this method, for a given $l$, $u$ is written 
as\footnote{This relation was first shown by \citet{legendre1769}. Of course, this 
was before Bessel functions were introduced, and the notation was different then.} 
\begin{eqnarray}
	u = l + 2\sum_{n=1}^\infty {{1}\over{n}}J_n(ne) \sin nl.
\label{eq:40}
\end{eqnarray}
This expansion converges for any value of $e <1$.
With the help of this expansion, the trigonometric functions of the true anomaly 
$f$ are expressed in terms of the mean anomaly $l$\footnote{With the expansions 
for $J_n(x)$, both $\cos f$ and $\sin f$, as well as some other functions relevant 
with them, can be expressed in powers of $e$. Extensive tabulations of series 
expansions are available in \citet{cayley1861}.}:
\begin{eqnarray}
	\cos f 
	=
	-e + {{2(1-e^2)}\over{e}}\sum_{n=1}^\infty J_n(ne)\cos nl
\label{eq:41}
\end{eqnarray}
\begin{eqnarray}
	\sin f 
	=
	2\sqrt{1-e^2}\sum_{n=1}^\infty 
	J_n'(ne)\sin nl,
\label{eq:42}
\end{eqnarray}
where $J_n'(x)$ denotes ${\rm d}J_n(x)/{\rm d}x$.
Hence,
\begin{eqnarray}
	v_{\rm rad, 1}
	&=&
	-(2\pi G{\rm M}_\odot)^{1/3} 
	\left({{m_1}\over{{\rm M}_\odot}}\right)^{1/3} q (1+q)^{-2/3} 
	P_{\rm orb}^{-1/3}
	\sin i
	\nonumber \\
	& & 
	\times
	\,\left[
	{{2\sqrt{1-e^2}}\over{e}}\sum_{n=1}^\infty J_n(ne)\cos nl
	\cos\varpi
	\right.
	\nonumber \\
	&&
	\left.
	-
	\sum_{n=1}^\infty 2
	J_n'(ne)\sin nl
	\sin\varpi
	\right] .
\label{eq:43}
\end{eqnarray}
Note that, although $e$ appears in the denominator, $e=0$ is not singular because 
$J_n(ne)$ reaches zero faster than $e$ itself as $e \rightarrow 0$. 

\begin{figure}
\centering
\includegraphics[width=1.0\linewidth,angle=0]{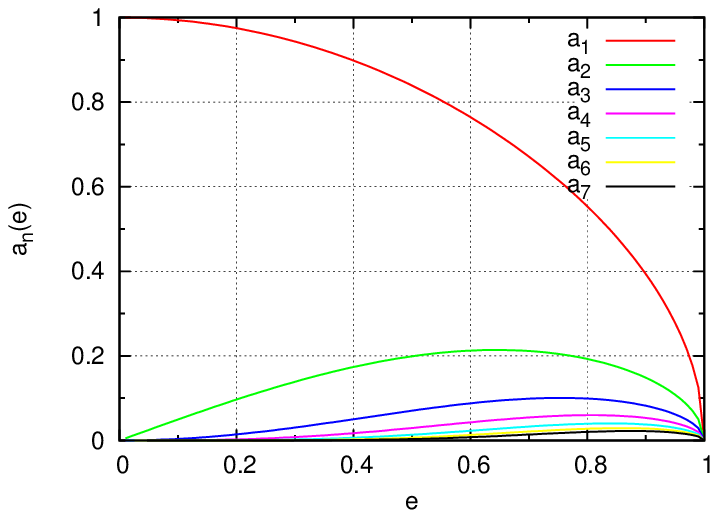}
\includegraphics[width=1.0\linewidth,angle=0]{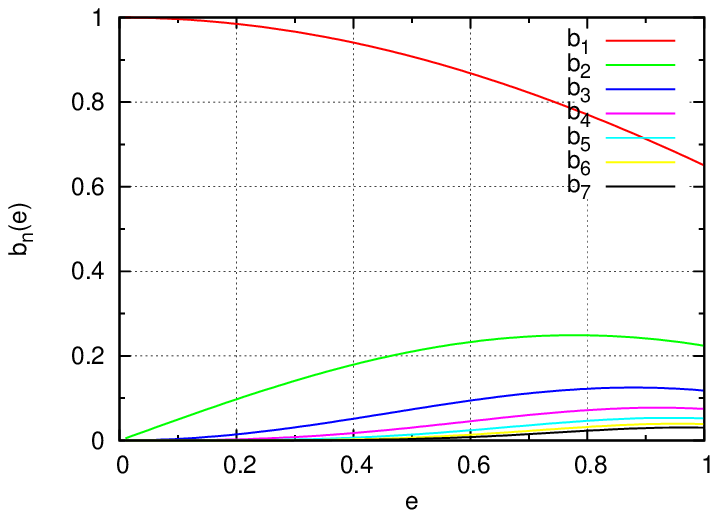}
\caption{Top panel: $a_n = 2e^{-1}(1-e^2)^{1/2}n^{-1}J_n(ne)$ as a function of 
$e$. Bottom panel: $b_n = 2n^{-1} J_n'(ne)$ as a function of $e$. }
\label{fig:11}
\end{figure}

Introducing
\begin{eqnarray}
	a_n(e) 
	\equiv
	{{2\sqrt{1-e^2}}\over{e}} {{1}\over{n}} J_n(ne)  
\label{eq:44}
\end{eqnarray}
and
\begin{eqnarray}
	b_n(e)
	\equiv
	{{2}\over{n}} 
	J_n'(ne) ,
\label{eq:45}
\end{eqnarray}
and with the help of equation\,(\ref{eq:39}),
we eventually obtain
\begin{eqnarray}
	&&
	{{\omega_0}\over{c}}\int_{t_0}^t 
	v_{\rm rad, 1}
	\,{\rm d}t'
	\nonumber \\
	& &
	=
	\alpha
	\left\{ \sum_{n=1}^\infty \xi_n(e,\varpi) 
	\sin \left[ n\Omega (t-t_0) + \vartheta_n(e,\varpi) \right] 
	\right.
	\nonumber \\
	& & 
	+\tau(e,\varpi) \bigg\},
\label{eq:46}
\end{eqnarray}
where
\begin{eqnarray}
	\xi_n (e,\varpi)
	= \sqrt{a_n^2 \cos^2\varpi + b_n^2 \sin^2\varpi},
\label{eq:47}
\end{eqnarray}
\begin{eqnarray}
	\vartheta_n (e,\varpi)=\tan^{-1} \left({{b_n}\over{a_n}} \tan\varpi\right)
\label{eq:48}
\end{eqnarray}
and
\begin{eqnarray}
	\tau(e,\varpi) = - \sum_{n=1}^\infty b_n\sin\varpi,
\label{eq:49}
\end{eqnarray}
and $\alpha$ is defined by equation\,(\ref{eq:10}).
Fig.\,\ref{fig:11} shows $a_n(e)$ and $b_n(e)$ as 
functions of $e$.

The difference in phase modulation between the circular orbit and ellipsoidal ones 
is that the former is expressed by a single angular frequency $\Omega$ while the latter is 
composed of harmonics of $\Omega$. This is of course naturally expected. In the 
series expansion in equation\,(\ref{eq:46}), the terms of $n=1$ dominate over the 
higher-order terms, but, in the case of $e \gtrsim 0.5$ and $\varpi \sim \pi/2$, 
the contribution of the higher-order terms becomes non-negligible. 

\subsubsection{In the limiting cases of $e=0$}
\label{sec:3.2.2}

One might worry whether the series-expansion form given in the above 
formally tends to the results obtained in the circular orbital case 
in the limit of $e \rightarrow 0$. In this section, we prove that it does. 
As $e \rightarrow 0$, equation (\ref{eq:36}) is reduced to
\begin{equation}
	v_{\rm rad, 1} = -\Omega a_1 \cos(f+\varpi) \sin i.
\label{eq:50}
\end{equation}
It should be remembered that in the case of $e=0$, 
the angle $\varpi$ can be arbitrarily chosen, 
while $(f+\varpi)$ is uniquely defined as the angle between 
the ascending node and the star at the moment. In the cases of $e=0$, 
$f=u=l=\Omega(t-t_0)$. Hence 
\begin{eqnarray}
	v_{\rm rad, 1} &=& -\Omega a_1 \cos[\Omega(t-t_0)+\varpi] \sin i
\nonumber\\
	&=&
	\Omega a_1 \cos[\Omega(t-t_0)+(\varpi+\pi)] \sin i.
\label{eq:51}
\end{eqnarray}
If we choose $\varpi=0$, $(t-t_0)$ denotes the time when the star crosses 
the descending node. As we defined in section \ref{sec:2.1}, 
$t=0$ is the time that the radial velocity reaches its maximum. 
Therefore, we set 
\begin{equation}
	\lim_{e\to 0}\varpi=0
\label{eq:52}
\end{equation}
and $t_0=0$ to reduce the above expression to the form given in section 
\ref{sec:2}.  

The apparently complex expansion in equation (\ref{eq:43}) also reverts 
to the above form. We only have to note that 
${\displaystyle\lim_{e\to 0} J_1'(e)=1/2}$
and 
${\displaystyle\lim_{e\to 0} J_1(e)/e =1/2}$,
while 
${\displaystyle\lim_{e\to 0} J_n'(ne)=0}$
and 
${\displaystyle\lim_{e\to 0} J_n(ne)/e =0}$
for $n\ge 2$. 

As for $a_n(e)$ and $b_n(e)$, as seen in Fig.\,\ref{fig:11}, 
$a_1(0)=1$
and 
$b_1(0)=1$,
while 
$a_n(0)=0$
and
$b_n(0)=0$
for $n\ge 2$.  
Hence,
\begin{equation}
	\lim_{e\to 0} \xi_1(e, \varpi) =1,
\label{eq:53}
\end{equation} 
\begin{equation}
	\lim_{e\to 0} \vartheta_1(e, \varpi) = 0,
\label{eq:54}
\end{equation}
and
\begin{equation}
	\lim_{e\to 0} \tau(e, \varpi) = 0,
\label{eq:55}
\end{equation}
while 
\begin{equation}
	\lim_{e\to 0} \xi_n(e, \varpi) =0
\label{eq:56}
\end{equation} 
and 
\begin{equation}
	\lim_{e\to 0} \vartheta_n(e, \varpi) = 0
\label{eq:57}
\end{equation}
for $n\ge 2$.
Therefore, as expected, with $e\rightarrow 0$, equation (\ref{eq:46}) tends to
\begin{eqnarray}
	{{\omega_0}\over{c}} \int_0^t v_{\rm rad, 1}\, {\rm d}t'
	=
	\alpha \sin \Omega t,
\label{eq:58}
\end{eqnarray}
which is identical to the second term in square brackets 
on the right-hand side of equation (\ref{eq:4}) for circular orbital motion.

\subsection{The expected frequency spectrum}
\label{sec:3.3}

\subsubsection{Mathematical formula}
\label{sec:3.3.1}

Although equation\,(\ref{eq:46}) is an infinite series expansion, in practice 
high-order components are negligibly small and we may truncate the expansion with 
certain finite terms. Indeed, Fig.\,\ref{fig:11} implies that this is 
true. In carrying out the Fourier analysis of pulsating stars showing phase 
modulation due to such orbital motion in a binary, the problem then becomes how to 
treat the terms $\cos[(\omega_0 t + \phi +\alpha\tau)  + \alpha\sum \xi_n\sin 
(n\Omega t + \vartheta_n)]$.  
Note that $\alpha\tau$ is constant for a given binary system and 
is common to all the harmonic components. It is not distinguishable 
from the intrinsic phase $\phi$, hence, hereafter, we adopt a new symbol
$\varphi \equiv \phi + \alpha\tau$ to represent a constant phase. 

With multiple and repetitive use  of the Jacobi-Anger expansion 
\citep{lebrun1977},
we get 
\begin{eqnarray}
&&
	\cos
	\left[ (\omega_0 t + \varphi)
	+
	\alpha \sum_{n=1}^N \xi_n \sin(n\Omega t + \vartheta_n) 
	\right]
	\nonumber \\
	&&
	=
	\sum_{k_1=-\infty}^\infty J_{k_1}(\alpha\xi_1)
	\nonumber\\
	&&
	\times
	\cos\left[ \omega_0 t + \varphi + \sum_{n=2}^N \xi_n\sin(n\Omega t + 
\vartheta_n)
	+k_1(\Omega t +\vartheta_1)\right]
	\nonumber \\
	&&
	=
	\sum_{k_1=-\infty}^\infty J_{k_1}(\alpha\xi_1)
	\sum_{k_2=-\infty}^\infty J_{k_2}(\alpha\xi_2)
	\nonumber\\
	&&
	\times
	\cos\bigg[ \omega_0 t + \varphi +k_1(\Omega t +\vartheta_1)
	+k_2(2\Omega t + \vartheta_2)
	\nonumber\\
	&&
	\left.	
	+\sum_{n=3}^N \xi_n\sin(n\Omega t + \vartheta_n)
	\right]
	\nonumber\\
	&&
	=
	\cdots\cdots
	\nonumber\\
	&&
	=
	\sum_{k_1=-\infty}^\infty \cdots\sum_{k_N=-\infty}^\infty
	\left[
	\prod_{n=1}^N J_{k_n}(\alpha \xi_n) 
	\right]
	\nonumber\\
	&&
	\times
	\cos
	\left[ \omega_0 t + \varphi
	+
	\sum_{n=1}^N k_n(n\Omega t + \vartheta_n) 
	\right] ,
\label{eq:59}
\end{eqnarray}
where $N$ denotes a large number with which the infinite series are truncated.
This is the most general formula, except for the truncation, covering the case of 
a circular orbit, which has already been discussed in the previous section.

\subsubsection{General description}
\label{sec:3.3.2}

The above result means that (i) whatever the pulsation mode is, the frequency 
spectrum shows a multiplet with each of adjacent components separated by the 
orbital frequency $\Omega$, (ii) the amplitude of these multiplet components is 
strongly dependent on  $\alpha$, which is defined by equation\,(\ref{eq:10}) and 
sensitive to the eccentricity, $e$, and the angle between the ascending node and 
the periapsis, $\varpi$, (iii) while the multiplet is symmetric in the case of a 
circular orbit, with increasing deviation from a circular orbit it becomes more 
asymmetric, (iv) while in the case of $\alpha \lesssim 1$ the multiplet is likely 
to be 
seen as a triplet for which the central component is the highest, in the case of 
$\alpha \gtrsim 1$ the multiplet will be observed as a quintuplet or higher-order 
multiplet and the side peaks will be higher than the central peak.

\subsubsection{The case of $\alpha \ll 1$}
\label{sec:3.3.3}

In the case of $\alpha \ll 1$, we may truncate the infinite series with $N=2$:
\begin{eqnarray}
	&&
	\cos
	\left[ (\omega_0 t + \varphi)
	+
	\alpha \sum_{n=1}^2 \xi_n \sin(n\Omega t + \vartheta_n)
	\right]
	\nonumber \\
	&=&
	\sum_{k_1=-\infty}^\infty \sum_{k_2=-\infty}^\infty
	J_{k_1}(\alpha \xi_1) J_{k_1}(\alpha \xi_2)
	\nonumber\\
	&&
	\times
	\cos
	\left[ \omega_0 t 
	+
	(k_1+2k_2)\Omega t 
	+ \varphi 
	+ k_1\vartheta_1
	+ k_2\vartheta_2
	\right]
	\nonumber \\
	&\simeq&
	J_0(\alpha \xi_1)J_0(\alpha \xi_2) \cos(\omega_0 t + \varphi)
	\nonumber \\
	& & 
	+ J_1(\alpha \xi_1)J_0(\alpha \xi_2) \cos[(\omega_0 + \Omega)t 
	+\varphi + \vartheta_1]
	\nonumber \\
	& & 
	+ J_{-1}(\alpha \xi_1)J_0(\alpha \xi_2) \cos[(\omega_0 
	- \Omega)t +\varphi - \vartheta_1]
	\nonumber \\
	& &
	+ J_0(\alpha \xi_1)J_1(\alpha \xi_2) \cos[(\omega_0 + 2\Omega)t 
	+ \varphi + \vartheta_2]
	\nonumber \\
	& & 
	+ J_0(\alpha \xi_1)J_{-1}(\alpha \xi_2) \cos[(\omega_0 - 2\Omega)t 
	+ \varphi - \vartheta_2]
	\nonumber \\
	& & 
	+ O(\alpha^2).
\label{eq:60}
\end{eqnarray}
From this, the amplitude ratios are derived as follows:
\begin{eqnarray}
	{{A_{+1}+A_{-1}}\over{A_0}} 
	&=& 
	{{2J_1(\alpha \xi_1)}\over{J_0(\alpha \xi_1)}}
\nonumber\\
	&\simeq& 
	\alpha \xi_1
\label{eq:61}
\end{eqnarray}
and
\begin{eqnarray}
	{{A_{+2}+A_{-2}}\over{A_0}} 
	&=& 
	{{2 J_1(\alpha \xi_2)}\over{J_0(\alpha \xi_2)}} 
\nonumber\\
	&\simeq& 
	\alpha \xi_2.
\label{eq:62}
\end{eqnarray}
Also, the following phase relations are derived:
\begin{eqnarray}
	{{\phi_{+1} - \phi_{-1}}\over{2}} = \vartheta_1
\label{eq:63}
\end{eqnarray}
and
\begin{eqnarray}
	{{\phi_{+2} - \phi_{-2}}\over{2}} = \vartheta_2.
\label{eq:64}
\end{eqnarray}
Note that $\xi_n$ and $\vartheta_n$ ($n=1,2$) are functions of $e$ and $\varpi$.
This means that the four constraints -- equations\,(\ref{eq:61}), (\ref{eq:62}), 
(\ref{eq:63}) and (\ref{eq:64}) -- are obtained for three quantities, $\alpha$, 
$e$, and $\varpi$. 

It should also be noted that 
\begin{eqnarray}
	{{A_{+1}-A_{-1}}\over{A_0}} = 0
\label{eq:65}
\end{eqnarray}
and 
\begin{eqnarray}
	{{A_{+2}-A_{-2}}\over{A_0}} = 0.
\label{eq:66}
\end{eqnarray}
Hence, the multiplet is symmetric with respect to the highest central peak.

\subsection{Procedures to derive binary parameters from the frequency spectrum}
\label{sec:3.4}

The three unknowns $\alpha$, $e$ and $\varpi$ can be determined from the frequency 
spectrum. To illustrate more clearly this new technique, we describe practical 
procedures for the case of $\alpha\ll 1$.

\subsubsection{Series expansion in terms of $e$}
\label{sec:3.4.1}

With use of the series-expansion form of the Bessel function, 
equation\,(\ref{eq:15}),
to the order of $O(e^6)$, the coefficients $a_1(e)$ and $b_1(e)$ are given by
\begin{eqnarray}
	a_1 (e)\simeq \sqrt{1-e^2} \left(1-{{1}\over{8}}e^2+{{1}\over{192}}e^4
	-{{1}\over{9216}}e^6\right)
\label{eq:67}
\end{eqnarray}
and
\begin{eqnarray}
	b_1 (e)\simeq 1-{{3}\over{8}}e^2+{{5}\over{192}}e^4-{{7}\over{9216}}e^6,
\label{eq:68}
\end{eqnarray}
respectively.
Also,
\begin{eqnarray}
	a_2 (e)\simeq {{e}\over{2}}\sqrt{1-e^2}
	\left(1-{{1}\over{3}}e^2+{{1}\over{24}}e^4-{{1}\over{360}}r^6\right)
\label{eq:69}
\end{eqnarray}
and
\begin{eqnarray}
	b_2 (e)\simeq {{e}\over{2}}
	\left(1-{{2}\over{3}}e^2+{{1}\over{8}}e^4-{{1}\over{90}}e^6\right).
\label{eq:70}
\end{eqnarray}
Substitution of these into equation\,(\ref{eq:47}) leads to explicit expressions 
for $\xi_1(e,\varpi)$ and $\xi_2(e,\varpi)$ with a series expansion of $e$. 
Combining these expressions with equations (\ref{eq:61}) and (\ref{eq:62}), 
we get
\begin{eqnarray}
	\left( {{A_{+1}+A_{-1}}\over{A_0}} \right)^2
	&=&
	\alpha^2
	\left[
	1-\left({{3}\over{4}}+{{1}\over{2}}\cos^2\varpi\right)e^2
	\right.
\nonumber\\
	& &
	+\left({{37}\over{192}}+{{1}\over{12}}\cos^2\varpi\right)e^4
\nonumber\\
	& &
	\left.
	-\left({{97}\over{4608}}+{{15}\over{2304}}\cos^2\varpi\right)e^6
	\right]
\label{eq:71}
\end{eqnarray}
and
\begin{eqnarray}
	\left( {{A_{+2}+A_{-2}}\over{A_0}} \right)^2
	&=&
	{{\alpha^2}\over{4}}e^2
	\left[
	1-\left( {{4}\over{3}}+{{1}\over{3}}\cos^2\varpi \right)e^2
	\right.
\nonumber\\
	& &
	\left.
	+\left({{25}\over{36}}+{{1}\over{6}}\cos^2\varpi\right)e^4	
	\right].
\label{eq:72}
\end{eqnarray}
Similarly, from equations (\ref{eq:48}), $\vartheta_n (e,\varpi)$ $(n=1,2)$ 
is explicitly written with a series expansion of $e$, and by combining them 
with the phase differences of the sidelobes, we get
\begin{eqnarray}
&&
	\tan{{\phi_{+1}-\phi_{-1}}\over{2}}
\nonumber\\
	&=&
	{{1}\over{\sqrt{1-e^2}}}
	\left( 1-{{1}\over{4}}e^2-{{1}\over{96}}e^4-{{1}\over{1536}}e^6 \right)
	\tan\varpi
\label{eq:73}
\end{eqnarray}
and
\begin{eqnarray}
&&
	\tan{{\phi_{+2}-\phi_{-2}}\over{2}}
\nonumber\\
	&=&
	{{1}\over{\sqrt{1-e^2}}}
	\left( 1-{{1}\over{3}}e^2-{{1}\over{36}}e^4-{{1}\over{270}}e^6 \right)
	\tan\varpi	 .
\label{eq:74}
\end{eqnarray}

\subsubsection{Procedures to determine the binary parameters}
\label{sec:3.4.2}

The left-hand sides of equations (\ref{eq:71}) -- (\ref{eq:74}) are observables, 
thus the three unknowns, $\alpha$, $e$ and $\varpi$, can be derived from these 
equations. Since the number of unknowns is smaller than the number of equations, 
the solution is not uniquely determined. Solutions satisfying all the constraints 
within the observational errors should be sought. 

\begin{figure}
\centering
\includegraphics[width=1.0\linewidth,angle=0]{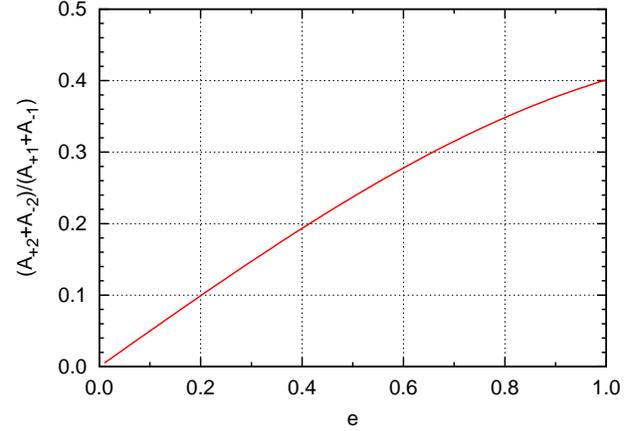}
\caption{The expected amplitude ratio of the second sidelobes to the first 
sidelobes as a function of eccentricity $e$, in the case of $\varpi=0$ and 
$\alpha \ll 1$.}
\label{fig:12}
\end{figure}

Note that, as seen in equations\,(\ref{eq:71}) -- (\ref{eq:74}), only the 
constraint $(A_{+2}+A_{-2})/A_0$ among the four constraints is of the order of 
$O(e^1)$. A reasonably good estimate of the eccentricity $e$ can be deduced 
from this. Given that, 
\begin{eqnarray}
	{{\xi_2}\over{\xi_1}} 
	={{J_2(2e)}\over{2J_1(e)}}
	\simeq {{e}\over{2}},
\label{eq:75}
\end{eqnarray}
we can derive the eccentricity $e$:
\begin{eqnarray}
	e \simeq  {{2(A_{+2}+A_{-2})}\over{A_{+1}+A_{-1}}}.
\label{eq:76}
\end{eqnarray}
Fig.\,\ref{fig:12} demonstrates that this is a good approximation. By substituting 
the value of $e$ estimated in this way into equation\,(\ref{eq:73}), we get a 
first guess for $\varpi$. Then by putting the estimated values of $e$ and $\varpi$ 
into equation\,(\ref{eq:71}), we get the value of $\alpha$. Better solutions are 
obtained by iteration. Once the value of $\alpha$ is derived, the mass function is 
determined by equation\,(\ref{eq:22}), and the projection of the semi-major axis 
of the orbit into the celestial plane, $a_1\sin i$, is deduced to be
\begin{eqnarray}
	a_1\sin i={{P_{\rm osc}}\over{2\pi}}\alpha c.
\label{eq:77}
\end{eqnarray}

The radial velocity is approximately determined with the four terms: $a_n$ and 
$b_n$ for $n=1$ and $2$:
\begin{eqnarray}
	v_{\rm rad, 1} (t) 
	&=& {{P_{\rm osc}}\over{P_{\rm orb}}} \alpha c
	\nonumber\\
	&&
	\times\left[(a_1(e)\cos\Omega t+2a_2(e)\cos 2\Omega t)\cos\varpi\right.
	\nonumber\\
	&&
	-\left. (b_1(e)\sin\Omega t + 2b_2(e)\sin 2\Omega t)\sin\varpi\right],
\label{eq:78}
\end{eqnarray}
where $a_1(e)$, $a_2(e)$, $b_1(e)$ and $b_2(e)$ are given by 
equations\,(\ref{eq:67})--(\ref{eq:70}).

As seen in Fig.\,\ref{fig:11}, $a_3(e)$, $b_3(e)$ and higher-order terms are 
negligibly small up to $e \lesssim 0.2$. So the radial velocity derived in this 
way is acceptable in the case of $e \lesssim 0.2$. Higher-order terms become 
important with increasing $e$. Those terms are available from the third and 
higher-order sidelobes.

\subsubsection{The case of $e \ll 1$}
\label{sec:3.4.3}

In the case of $e \ll 1$, as seen in equation\,(\ref{eq:71}),
$(A_{+1}+A_{-1})/A_0 \simeq \alpha \times (1+O(e^2))$.
Hence, to the order of $O(e^1)$, the eccentricity is determined by 
equation\,(\ref{eq:76}), 
and $\alpha$ is determined by
\begin{eqnarray}
	\alpha \simeq 
	{{A_{+1}+A_{-1}}\over{A_0}}.
\label{eq:79}
\end{eqnarray}
Similarly, within the same approximation, from equation\,(\ref{eq:73}),
\begin{eqnarray}
	\varpi \simeq {{\phi_{+1}-\phi_{-1}}\over{2}}.
\label{eq:80}
\end{eqnarray}
As discussed in section \ref{sec:3.2.2}, in the limit of 
$e\rightarrow 0$, $\varpi$ tends to $0$. Then, 
in this limit, $\phi_{+1}$ and $\phi_{-1}$ become equal each other, 
as expected from the analysis of circular orbits.

The mass function is given by
\begin{eqnarray}
	f(m_1,m_2,\sin i) 
	=
	\left(\frac{A_{+1}+A_{-1}}{A_0}\right)^3 
	\frac{P_{\rm osc}^{3}}{P_{\rm orb}^{2}}
 	\frac{c^3}{2 \pi G},
\label{eq:81}
\end{eqnarray}
and $a_1\sin i$ is given by
\begin{eqnarray}
	a_1\sin i \simeq {{P_{\rm osc}}\over{2\pi}}{{A_{+1}+A_{-1}}\over{A_0}} c.
\label{eq:82}
\end{eqnarray}
The radial velocity is obtained by setting $a_1(e)=b_1(e)=1$ and 
$a_2(e)=b_2(e)\simeq e/2$:
\begin{eqnarray}
	v_{\rm rad, 1} (t) 
	&\simeq& {{P_{\rm osc}}\over{P_{\rm orb}}} {{A_{+1}+A_{-1}}\over{A_0}} c
	\nonumber\\
	&&
	\times\left[(\cos\Omega t+e\cos 2\Omega t)
	\cos\left({{\phi_{+1}-\phi_{-1}}\over{2}}\right)\right.
	\nonumber\\
	&&
	-\left. (\sin\Omega t + e\sin 2\Omega t)
	\sin\left({{\phi_{+1}-\phi_{-1}}\over{2}}\right)\right].
\label{eq:83}
\end{eqnarray}

\subsection{Some more examples with artificial data}
\label{sec:3.5}

\subsubsection{An example for the case of $\alpha \ll 1$}
\label{sec:3.5.1}

In order to see how the present method works, we generate artificial, noise-free 
light curve data. The input parameters are: $m_1 = m_2 = 2$\,M$_{\odot}$, $e = 
0.3$, $\varpi = 0$, $i=90^\circ$, 
$\nu_{\rm osc} (\equiv 1/P_{\rm osc}) = 20$\,d$^{-1}$ and 
$P_{\rm orb} = 1$\,d, giving $\alpha = 1.13 \times 10^{-2}$. This case could apply 
to a binary star with two $\delta$\,Sct stars in an eccentric orbit. The infinite 
series of $a_n$ and $b_n$ were truncated at $N=150$.

\begin{figure}
\centering
\includegraphics[width=1.0\linewidth,angle=0]{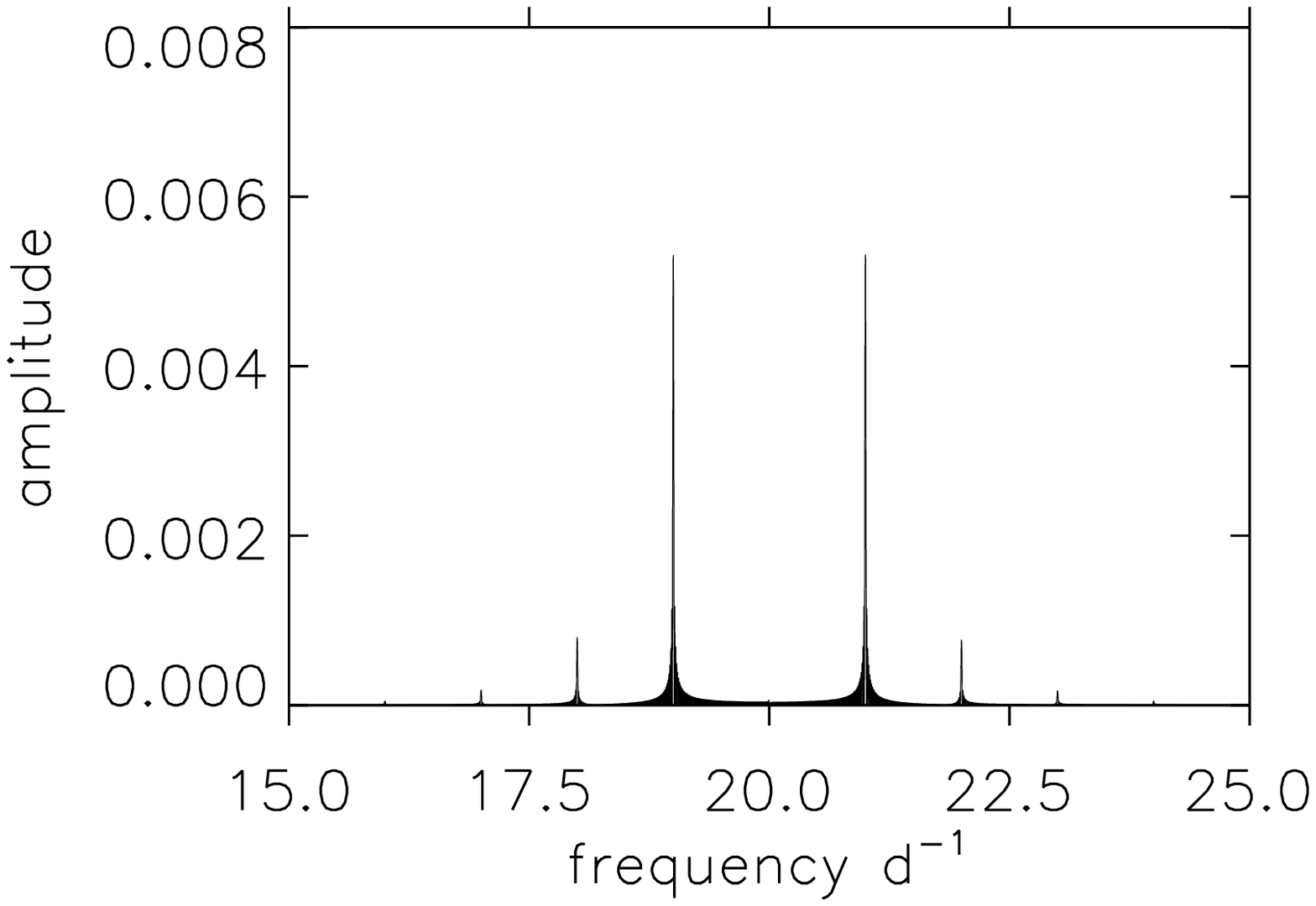}
\includegraphics[width=1.0\linewidth,angle=0]{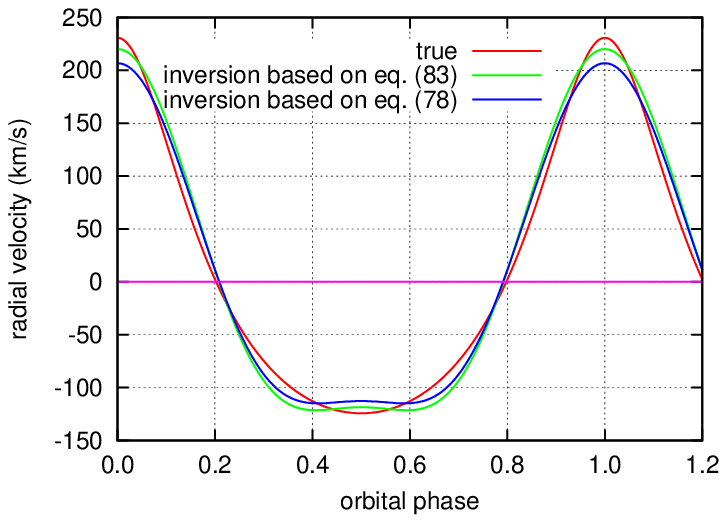}
\caption{Top panel: An amplitude spectrum for artificial data with $m_1 = m_2 = 
2$\,M$_{\odot}$, $e = 0.3$, $\varpi = 0$, $i=90^\circ$, 
$\nu_{\rm osc} = 20$\,d$^{-1}$ and 
$P_{\rm orb} = 1$\,d after prewhitening the central peak, which has an intensity 
amplitude of 1.0, by definition. Bottom panel: The true radial velocity curve 
(red) and two photometrically determined ones (green and blue). The green curve is 
obtained from the first and the second sidelobes and the central peak in the 
amplitude spectrum, based on equation\,(\ref{eq:83}); 
the blue curve is a solution based on equation\,(\ref{eq:78}). Note that 
they are obtained without iteration.}
\label{fig:13}
\end{figure}

The top panel of Fig.\,\ref{fig:13} shows the amplitude spectrum of the generated 
data sampled with 10 points per pulsation cycle over a time span of 10 orbital 
periods after the central peak of amplitude 1.0 (in intensity) has been 
prewhitened. There is no phase difference between the sidelobes. From this 
$\varpi=0$ is deduced. The first sidelobes have almost equal amplitudes of only 
$5.4\times 10^{-3}$ the amplitude of the central peak. This is consistent with 
equations (\ref{eq:62}) and (\ref{eq:65}). The second sidelobes also have almost 
equal amplitudes, and their amplitude ratio to the first sidelobes is 0.15, as 
expected from equation (\ref{eq:76}). Further sidelobes have such small amplitudes 
that they are unlikely to be observed in this case.  

From the amplitude ratio of the second sidelobes to the first sidelobes, $0.15$, 
the value of eccentricity $e=0.30$ is reproduced well. 
Combining this with the amplitude ratio 
of the first sidelobes to the central peak, $0.0108$, 
we determine $\alpha$ and the mass 
function; $1.08\times 10^{-2}$ and $0.43\,{\rm M}_\odot$, respectively.  
These values determined from the amplitude spectrum are also in satisfactory 
agreement with the true values; 
$1.13\times 10^{-2}$ and $0.5\,{\rm M}_\odot$, respectively. 

The bottom panel of Fig.\,\ref{fig:13} demonstrates how well the radial velocity 
curve is reproduced. The true radial velocity is shown with red, and the one 
determined from the amplitude spectrum is shown with green. The latter is the 
solution of $O(e^1)$ obtained from equation\,(\ref{eq:83}). This is the crudest 
solution. The blue curve corresponds to the radial velocity curve that can 
ultimately be determined within this framework from equation\,(\ref{eq:78}) with a 
higher-order approximation. We see the radial velocity curve is reasonably well 
reproduced.

To summarise this experiment with artificial data: We conclude that in the case of 
$\alpha \ll 1$ the binary parameters are reproduced well from the photometric 
light curve alone.

\subsubsection{A more extreme case of $\alpha >1$}
\label{sec:3.5.2}

\begin{figure}
\centering
\includegraphics[width=1.0\linewidth,angle=0]{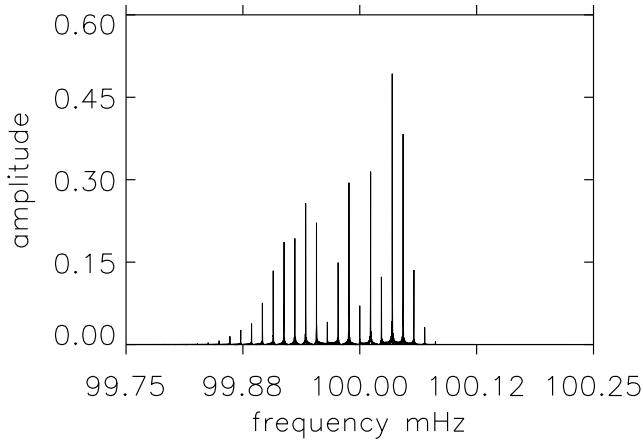}
\caption{The amplitude spectrum for artificial data with $m_1 = m_2 = 
3$\,M$_{\odot}$, $e = 0.3$, $\varpi = 0$, $i=90^\circ$, 
$\nu_{\rm osc} = 100$\,mHz ($P_{\rm osc} 
= 10$\,s) and $P_{\rm orb} = 1$\,d; $\alpha = 5.6$. The amplitude scale is in 
fractional intensity.}
\label{fig:14}
\end{figure}

We now note that amplitude spectra are not so simple in all cases. Taking an 
extreme example, Fig.\,\ref{fig:14} shows a case where $\alpha = 5.6$ that is 
equivalent to a 10-s pulsar in a 1-d binary with another neutron star. The 
parameters are: $m_1 = m_2 = 3$\,M$_{\odot}$, $e = 0.3$, $\varpi = 0$, $\nu_{\rm 
osc} = 100$\,mHz, $P_{\rm orb} = 1$\,d and $i = 90^\circ$. Note that the central 
peak of the multiplet in this case has almost no amplitude, and that the pattern 
is highly asymmetric. 

\subsection{An actual example, KIC\,4150611 revisited}
\label{sec:3.6}

In section\,\ref{sec:2.6}, we assumed that KIC\,4150611 has a circular orbit. How 
is this assumption justified? In the case of an eccentric orbit, the amplitude 
ratio of the second sidelobes to the first sidelobes is proportional to $e$, as 
long as $\alpha \ll 1$. In the case of KIC\,4150611, the multiplet is seen as a 
triplet, not a quintuplet. Hence, the eccentricity is smaller than the detection 
limit, and the assumption of a circular orbit adopted in section\,\ref{sec:2.6} is 
consistent with the observation. 

From Table\,\ref{table:3} we see that the root-mean-square error on amplitude is 
0.007\,mmag, hence at the 1$\sigma$ level an upper limit to the eccentricity can 
be made from the noise level for the quintuplet sidelobes. From equation\,(68) we 
then find $e \le 0.12$ for the highest amplitude pulsation frequency. Hence at the 
3$\sigma$ level the constraint on the eccentricity is weak. Supposing that the 
orbit of KIC\,4150611 is non-circular, we estimate the angle $\varpi$ from the 
phase difference $(\phi_{+1}-\phi_{-1})/2$. This value is $-0.02 \pm 0.04$ radians 
for the triplet of $\nu_{\rm osc}=22.619577$\,d$^{-1}$. The estimated value of 
$\alpha$ is the same as that determined with the assumption of a circular orbit, 
thus the mass function is the same as that derived in section\,\ref{sec:2.6}.

%%%%%%%%%%%%%%%%%%%%%%%%%%%%%%%%%%%%%%
\section{Exoplanet hunting}
\label{sec:4}

A new application of our method is in the search for exoplanets. The prime goal of 
the CoRoT and {\it Kepler} missions is to find exoplanets by the transit method. 
The other main technique for exoplanet searches is the radial velocity technique 
using high-resolution ground-based spectroscopy. Both of these techniques have 
concentrated on solar-like and lower main sequence stars where planetary transits 
are deeper than for hotter main sequence stars, and where radial velocities are 
greater than for more massive host stars. Now with our new technique of 
photometric measurement of radial velocity, there is a possibility to detect 
planets around hotter main sequence pulsating stars, such as $\delta$\,Sct and 
$\beta$\,Cep stars, and around compact pulsating stars, such as subdwarf B 
pulsators and various pulsating white dwarf stars. This has been demonstrated in 
the case of the subdwarf B pulsator V391\,Peg \citep{silvottietal2007} using the 
$O-C$ method.

Let us look here at the possibility of exoplanet detection using our method. It 
can be seen in equations\,(\ref{eq:21}) and (\ref{eq:25}) 
that the frequency triplet in the amplitude 
spectrum of a pulsating star with a planetary companion will have larger sidelobes 
for higher pulsation frequency and for longer orbital period. We therefore examine 
a limiting case for the {\it Kepler} mission of a 1.7-M$_{\odot}$ $\delta$\,Sct 
star with a pulsation frequency of 50\,d$^{-1}$ and with a planetary companion of 
one Jupiter mass ($10^{-3}$\,M$_{\odot}$), with an orbital period of 300\,d and an 
inclination of $i=90^\circ$. In this case $\alpha = 1.1 \times 10^{-3}$ is small. 

We find that the first orbital sidelobes have amplitudes of $558 \times 10^{-6}$ 
where the central peak has an amplitude of 1. With {\it Kepler} data we can reach 
photometric precision of a few $\mu$mag, so can detect signals of, say, 
10\,$\mu$mag and higher. This means that we need to have pulsation amplitudes of 
0.02\,mag, or more, to detect a Jupiter-mass planet with the orbital parameters 
given above. This is possible; there are, for example, $\delta$\,Sct stars with 
amplitudes greater than 0.02\,mag. 

To take another case, can we detect a hot 10-M$_{\rm Jupiter}$ planet in a 10-d 
orbit around a $\delta$\,Sct star? In this case $\alpha = 1.2 \times 10^{-3}$, so 
the detection limit is about the same as above: the sidelobes have amplitudes of 
$576 \times 10^{-6}$ where the central peak has an amplitude of 1. Signals such as 
these should be searched for in High Amplitude $\delta$\,Sct (HADS) stars. Compact 
stars also offer potential exoplanet discoveries with our technique.

Through photometric radial velocity measurement, the mass of the exoplanet can 
then be estimated with an assumption about the mass of the host star. If planetary 
transits are detected for the same exoplanet system, the size of the planet can 
also be derived. Hence the mean density of the exoplanet can be determined only 
through the photometric observations.

%%%%%%%%%%%%%%%%%%%%%%%%%%%%%%%%%%%%%%
\section{Discussion}
\label{sec:5}

\subsection{Photometric radial velocity measurement}
\label{sec:5.1}

As clearly seen in equation\,(\ref{eq:1}), the phase of the luminosity variation 
of a pulsating star in a binary system has information about the radial velocity 
due to the orbital motion. By taking the time derivative of the phase of 
pulsation, we can obtain the radial velocity at each phase of the orbital motion. 
We have shown in this paper that the Fourier transform of the light curve of such 
a pulsating star leads to frequency multiplets in the amplitude spectra where the 
frequency splitting and the amplitudes and phases of the components of the 
frequency multiplet can be used to derive all of the information traditionally 
found from radial velocity curves.

This is a new way of measuring the radial velocity. Until now, to measure radial 
velocity we have had to carry out spectroscopic observations of the Doppler shift 
of spectral lines\footnote{With the recent exception of the determination of 
radial velocity amplitude using `Doppler boosting', as in the example of the 
subdwarf B star -- white dwarf binary KPD\,1946+4340 using {\it Kepler} data 
\citep{bloemenetal2011}.}. In contrast, the present result means that radial 
velocity can be obtained from photometric observations alone. In the case of 
conventional ground-based observations, getting precise, uninterrupted 
measurements of luminosity variations is highly challenging at mmag precision. 
However, this situation has changed dramatically with space missions such as CoRoT 
and {\it Kepler}, which have $\mu$mag precision with duty cycles exceeding 
90\,per\,cent. Telescope time for spectroscopic observations is competitive, and 
suffers from most of the same ground-based limitations as photometry. Now for 
binary stars with pulsating components, a full orbital solution can be obtained 
from the light curves alone using the theory we have presented. 

\subsection{Limitations}
\label{sec:5.2}

There are clear limitations to the application of the theory presented here. 
Firstly, the pulsating stars to be studied must have stable pulsation frequencies. 
Many pulsating stars do not. In the Sun, for example, the pulsation frequencies 
vary with the solar cycle. That would not be much limitation for current {\it 
Kepler} data, since the solar cycle period is so long compared with the time span 
of the data. But other pulsating stars also show frequency variability, and on 
shorter time scales. RR\,Lyrae stars show the Blazhko effect, which has frequency, 
as well as amplitude variability. Other pulsating stars show frequency changes 
much larger than expected from evolution on time scales that are relevant here. 
Thus, to apply this new technique, a first step is to find pulsating stars with 
stable frequencies and binary companions. The best way to do this is to search 
for the frequency patterns we have illustrated in this work.

Another limitation can arise from amplitude modulation in a pulsating star. While 
this will not cause frequency shifts, it will generate a set of Fourier peaks in 
the amplitude spectrum that describe the amplitude modulation. For nonperiodic 
modulation on time scales comparable to the orbital period, the radial velocity 
sidelobe signal may be lost in the noise. 

Of course, it is imperative when frequency triplets or multiplets are found to 
distinguish among rotational multiplets with $m$\,modes, oblique pulsator 
multiplets that are pure amplitude modulation, and the frequency multiplets caused 
by frequency modulations. As we have explained in this paper, this can be done by 
careful examination of frequency separations, amplitudes and phases. For the 
latter, a correct choice of the time zero point is imperative. 

The important characteristics of FM multiplets is that the amplitude ratio of the 
sidelobes to the central peak is the same for all pulsation frequencies, and (for 
low eccentricity systems) the phases of the sidelobes are in quadrature with that 
of the central peak at the time of zero radial velocity, e.g., the time of eclipse 
for $i = 90^\circ$. We anticipate more of these stars being found and 
astrophysically exploited.

%%%%%%%%%%%%%%%%%%%%%%%%%%%%%%%%%%%%%%
\section*{acknowledgements}
This work was carried out with support from a Royal Society UK-Japan International 
Joint Program grant.

%%%%%%%%%%%%%%%%%%%%%%%%%%%%%%%%%%%%%%
\bibliography{fmbib2.bib}

\end{document}